\documentclass[
aip,
jcp,
 amsmath,amssymb,
 reprint,%
]{revtex4-1}

\usepackage{graphicx}
\usepackage{dcolumn}
\usepackage{bm}
\usepackage{amsmath,amsfonts,amssymb}
\usepackage{soul}

\usepackage{xcolor}
\usepackage{mathtools}
\usepackage{float}
\usepackage{lipsum}

\usepackage[utf8]{inputenc}
\usepackage[T1]{fontenc}
\usepackage{mathptmx}
\usepackage{etoolbox}

\makeatletter
\def\@email#1#2{%
 \endgroup
 \patchcmd{\titleblock@produce}
  {\frontmatter@RRAPformat}
  {\frontmatter@RRAPformat{\produce@RRAP{*#1\href{mailto:#2}{#2}}}\frontmatter@RRAPformat}
  {}{}
}%
\makeatother
\begin{document}

\preprint{AIP/123-QED}

\title[Polyamorphism in Glassy Network Materials]{Polyamorphism in Glassy Network Materials}
\author{M. H. Brown}
\affiliation{
Center for Theoretical Biological Physics, Rice University, Houston, Texas 77005, USA
}
\affiliation{ 
Department of Physics and Astronomy, Rice University, Houston, Texas 77005, USA
}

\author{P. G. Wolynes}%
 \email{pwolynes@rice.edu.}

\affiliation{
Center for Theoretical Biological Physics, Rice University, Houston, Texas 77005, USA
}
\affiliation{ 
Department of Physics and Astronomy, Rice University, Houston, Texas 77005, USA
}
\affiliation{
Department of Chemistry, Rice University, Houston, Texas 77005, USA
}

\date{\today}

\begin{abstract}
One dramatic feature of network liquids is the emergence at low temperatures and high pressures of polyamorphism, where multiple distinct liquid phases are accessed in a single material. Polyamorphism can arise from the competition between distinct local inherent structures corresponding to bonded and nonbonded ordering. Thermal bond breaking thus can lead to a phase transition often accompanied by thermodynamic anomalies away from the transition itself, such as the familiar density maximum in water at atmospheric pressure and $4^\circ$ C. Water exhibits network interactions in the form of hydrogen bonding between water molecules. The polyamorphic transition in water, however, is difficult to study due to the rapid crystallization of supercooled water and due to glassy effects at low temperatures. In the present work, we propose a simple microscopic model where the glassy and thermodynamic properties are both calculated directly from the microscopic potentials. The model contains a liquid-liquid phase transition, which, after tuning the microscopic parameters, may be located either above, near, or below the glass transition. By applying the Random First Order Transition theory of the glass transition to this simple microscopic model, we shine light on the interplay of polyamorphism and glassy properties in network liquids. We show the connection between the thermodynamic water-like anomalies and corresponding anomalies in the glassy kinetics. The analysis unveils key details on the way glassy dynamics modifies the phase transition kinetics. When the parameters of the model are tuned to produce a phase diagram resembling that of water, the liquid-liquid phase transformation near $T_g$ occurs via ``nanonucleation'', resulting in extremely small domains sizes and nonclassical nucleation kinetics which are predicted from the RFOT theory.
\end{abstract}

\maketitle
\section{\label{sec:intro}Introduction}
Network forming materials display a rich variety of macroscopic forms. Much of minerology's complexity, for example, arises from the many distinct crystalline forms, called polymorphs, of silicates. Technology relies on the many solid yet nonperiodic forms of glasses that arise when network liquids are rapidly cooled, avoiding the nucleation and growth of crystals. Unlike crystalline polymorphs, the detailed properties of such glasses depend on their history of preparation. Network formers in their fully liquid forms also display sometimes puzzling thermodynamic patterns that suggest they may be able to form structurally and thermodynamically distinct dense fluid phases. Such fluid phases have been called ``polyamorphs''. The most notorious example of such a possibility is water \cite{liu_liquid-liquid_2012, gallo_water_2016}. The well known density maximum of water at 4$^\circ$C early on encouraged the idea of there being two distinct forms of liquid water which could coexist or interconvert \cite{poole_phase_1992}, although other scenarios were also proposed \cite{speedy_stability-limit_1982, sastry_singularity-free_1996}. Extrapolating these data indicated that thermodynamic polyamorphism might occur in a deeply supercooled regime. Unfortunately, in the laboratory, crystallization is hard to avoid when liquid water is cooled to the suggested temperatures. Glassy amorphous solid forms of water are, however, easily accessed by other routes, both in the laboratory \cite{burton_crystal_1935, mishima_melting_1984} and the cosmos \cite{jenniskens_liquid_1997}. The relation of these glassy forms to a thermodynamic polyamorphic transition raises interesting questions, that, in this paper, we will try to address in the context of a tunable model of network materials that we introduced recently to describe network glasses within the random first order transition theory framework \cite{brown_bonding_2025}. Indeed, our own interest in the question of polyamorphism arose from our noting that there seemed to be modest deviations in the experimental data from our simple model predictions of the way the fragility and glass transition temperature vary with composition in doped silica and germania glasses. We ascribed these deviations, in part, to thermodynamic variations of the configurational entropy like those known in water. We note, of course, that since many glasses used in technology are mixtures, possibilities of distinct multiphase glass behavior abound, simply through compositional phase separation. This possibility was exploited in the invention of Vycor glass, decades ago \cite{watanabe_electron_1959}.

Both discrete polyamorphism of the type imputed to water and the glass transition are encouraged by the low configurational entropy of network materials which through frustration between different sources of stability allows distinct local structures to compete with each other \cite{tanaka_general_2000}. This propinquity between polyamorphs and glasses encouraged the frustration limited domain theory put forward by Kivelson et al. \cite{kivelson_thermodynamic_1995}. To a certain extent one can imagine the RFOT's random energy landscape, described globally by one-step replica symmetry breaking in the mean field limit \cite{kirkpatrick_connections_1987, kirkpatrick_scaling_1989}, as corresponding to the extreme limit of there being a near infinite, or at least very large, number of polyamorphs.

In this paper we will explore the interplay of a single discrete polyamorphic transition and the glass transition in a simple microscopic model of a one component network material. In addition to exploring the phase diagrams of the models, we will explore various regimes of nucleation mechanisms of the polyamorphs which are produced when they occur in the proximity to the glassy dynamics. As noted earlier by Stevenson and Wolynes \cite{stevenson_ultimate_2011}, the closeness of a first order transition to a glass transition allows discrete polyamorphs to interconvert through a non-classical mechanism called ``nanopercolative nucleation'' and we see this is true for the present model when it is tuned to resemble water.

The organization of the paper is as follows. We first propose in section \ref{sec: methods} a simple model for a fluid with two distinct liquid phases which may be analyzed both in terms of a liquid-liquid phase transition and in terms of the glass transition based on the same microscopic potentials. We review the approximation methods used to find the equilibrium liquid phase diagram, along with the physics governing the glass transition in the RFOT theory, where the behavior of individual glassy states is computed in the self-consistent phonon theory. In section \ref{sec: points of interest} several of the main features which characterize the behavior of the model, including the presence of liquid-gas and liquid-liquid phase transitions, water-like anomalies, and glass transitions at low temperatures are described. In section \ref{sec: investigation of model parameters}, we tune several of the key parameters of the model corresponding to the strengths of certain microscopic interactions in order to study how the phase diagram varies depending on the microscopic potentials. We then study the phase diagrams and glassy landmarks for several of these parameter choices in more detail in section \ref{sec: phase diagrams}, in order to see the behavior of three representative liquids with varying proximity between the liquid-liquid critical point and the glass transition. Last, in \ref{sec: nucleation}, we discuss nucleation processes for the transformation of one liquid to another within our model.

\section{Methods}\label{sec: methods}
\subsection{Model}\label{ssec: model}
We begin by considering a liquid made up of $N$ identical particles which may freely transform between two states: a ``bonding" state, and a ``non-bonding" state. These two states are analogous to the ``up" and ``down" spin states of the model of polyamorphs studied by Lee and Swendsen \cite{lee_simple_2001}. A bonding particle has a larger diameter $d_b$ than does a non-bonding particle which has a smaller diameter $d_u$ which allows the nonbonding particle to enter the solvation shell of a bonded molecule. At any moment in time, some fraction $x$ of the particles are in the bonding state, while the remaining fraction $1-x$ of the total particles are non-bonding. Macroscopically, the liquid phase is ultimately described by the fraction of bonding particles, $x$, the density of particles $\rho = N/V$ and the temperature $T$. In our previous similar model of network glass formers \cite{brown_bonding_2025}, the parameter $x$ was set by hand in order to describe mixtures of given composition, where the non-bonding and bonding particles represent different types of molecular units that do not interconvert and $x$ could be determined from the composition. Here, we instead compute the Gibbs free energy per particle $g(\rho, x; T, P,...)$ and allow the equilibrium values of $x$ to be determined naturally by minimizing $g$ with respect to $x$ (throughout this paper, extensive thermodynamic quantities are written with capital letters and the corresponding quantities ``per particle'' are written with lowercase letters). In this way, since $x$ is allowed to vary freely according to the microscopic dynamics of the liquid, we think of $x$ not as representing composition of different molecular units, but representing the relative proportion of different structural units in the liquid. Similar local structural order parameters have been used previously to investigate the glass transition \cite{charbonneau_dimensional_2012, tanaka_simple_1998} and polyamorphism \cite{chau_new_1998, liu_liquid-liquid_2012}. Additionally, at constant pressure minimizing $g$ with respect to $\rho$ determines the density. We will see later that $x$ is generally a robust order parameter for distinguishing the two liquid phases. Throughout most of the subcritical region of the phase diagram, in the high density liquid phase, $x\approx 1$ and nearly every particle is bonded, while in the low density liquid phase, $x\approx 0$ and nearly every particle is non-bonded.

In our model, a bonding particle interacts with other bonding particles with a bonding potential $V_{ij}=-\epsilon_b +\frac{1}{2}\kappa_b (r-d_b)^2$. This leads to a energy difference $\epsilon_b$, and an entropy difference $s_b = k_b\ln\left[\left(\frac{2\pi}{\beta \kappa_b}\right)^{1/2}\left(\frac{\rho}{e}\right)^{1/3}\right]$ between bonded particles and non-bonded particles, where $\beta = 1/k_BT$. In our numerical examples, we will take $\beta\kappa_b =300$ as the microscopic spring constant for bonds for most of the illustration. We have shown previously that this value is roughly representative of silica potentials \cite{hall_microscopic_2003, brown_bonding_2025}. 

In addition to forming bonds, pairs of particles are taken to interact via a Lennard-Jones potential $V_{ij}^{XY}=4\epsilon_{XY}\left[\left(\frac{\sigma_{XY}}{R_{ij}}\right)^{12}-\left(\frac{\sigma_{XY}}{R_{ij}}\right)^{6}\right]$, where $i$ and $j$ represent the particle identity and $X$ and $Y$ represent the bonding or non-bonding state of the $i$ and $j$ particles. For simplicity, we will assume additive bond diameters, $\sigma_{bb}$ = $d_b$, $\sigma_{uu} = d_u$, and $\sigma_{ub} = (d_b+d_u)/2$. In studying the model, we will allow ourselves to tune freely the parameters $\epsilon_b$, $\epsilon_{uu}$, $\epsilon_{bb}$, and $\epsilon_{ub}$ in order to explore the different behaviors of the liquid-liquid phase transition and their interplay with glassiness. The Lennard-Jones potential is split into a repulsive part
\begin{equation}
    v_{XY}(R)=\begin{cases}
        4\epsilon_{XY}\left[\left(\frac{\sigma_{XY}}{R}\right)^{12}-\left(\frac{\sigma_{XY}}{R}\right)^{6}\right]+\epsilon_{XY}, &R<2^{1/6}\sigma_{XY}\\
        0, & R\geq 2^{1/6}\sigma_{XY}
    \end{cases}
\end{equation}
and an attractive part
\begin{equation}
    u_{XY}(R)=\begin{cases}
        -\epsilon_{XY}, &R<2^{1/6}\sigma_{XY}\\
        4\epsilon_{XY}\left[\left(\frac{\sigma_{XY}}{R}\right)^{12}-\left(\frac{\sigma_{XY}}{R}\right)^{6}\right], & R\geq 2^{1/6}\sigma_{XY}
    \end{cases}
\end{equation}
This choice of separating the Lennard-Jones potential into attractive and repulsive parts based on the force is due to Weeks, Chandler, and Andersen \cite{andersen_relationship_1971}, although other choices \cite{kang_perturbation_1986} have been found to yield similar results and to be superior at the high densities that we often study here \cite{hall_intermolecular_2008}. The attractive part of the Lennard-Jones potential results in an effective van der Waals interaction with a van der Waals parameter $a_{XY}$ which is about $4.94\epsilon_{XY}d_u^3$ in value. In the free energy, one thus has an energetic term of the form $-\big[ a_{uu}(1-x)^2+a_{ub}2x(1-x)+a_{bb}x^2\big]\rho$ where $x$ is the fraction of particles in the bonding state. The repulsive part of the Lennard-Jones potential can be further approximated by an effective hard sphere repulsion with diameter\cite{barker_perturbation_1967} 
\begin{equation}\label{eq: Barker-Henderson}
    d_{HS}^{XY}=\int_0^{2^{1/6}\sigma_{XY}} \bigg(1-\exp(-v_{XY}(R)/k_BT)\bigg)dR.
\end{equation} 

A simple thermodynamic treatment of a mixture of hard spherical particles with different diameters was provided by Mansoori, Carnahan, Starling, and Leland \cite{mansoori_equilibrium_1971}. This treatment results in an interaction free energy analogous to that of the familiar Carnahan-Starling equation of state of the hard sphere fluid \cite{carnahan_equation_1969}. This equation of state shows excellent agreement with simulations at lower densities. We use a modified version of the Mansoori et al. form inspired by the equations of state of Wang, Khoshkbarchi, and Vera \cite{wang_new_1996}. This modification seeks to fix two issues with the Mansoori et al. equation of state seen at high densities: (a) the density at which the pressure diverges is too high, and (b), the power law of the divergence has the wrong exponent. According to the free volume theory of Salsburg and Wood \cite{salsburg_equation_1962}, the pressure should diverge as 
\begin{equation}
    P = k_B T\rho\frac{3\rho}{\rho_{div}-\rho}
\end{equation}
where $\rho_{div}$ is the maximum attainable density of the hard sphere particles, i.e. the density in a random close-packed state. We employ, then, an expansion of the form

\begin{gather}
    \frac{P}{\rho k_BT} = Z_{nb} =  \frac{\eta_{div}(x)}{\eta_{div}(x)-\eta}\sum_{i = 1}^\infty A_i(x) \eta^i\\
    \eta = \frac{\pi}{6}\rho\bigg[(1-x)(d_{HS}^{uu})^3+x(d_{HS}^{bb})^3\bigg].
\end{gather}
This functional form has the benefit of diverging with the correct power law exponent assuming a unique densest packing. We choose values for the $A_i$ to match the virial coefficients of the equation of state of Mansoori et al. \cite{mansoori_equilibrium_1971}. The expansion is truncated after 7 terms, although more terms could be used to similar effect. Note that if we were to set $\eta_{div} = 1$, we would obtain the Mansoori et al. expression exactly. Instead, to describe our high density system we use an empirical expression for the random close packing of mixtures due to Brouwers \cite{brouwers_packing_2007} $\eta_{div} = \eta_{RCP}\approx 0.64\cdot(1+4\cdot0.1306\cdot(1-x)\cdot x\cdot(d_b/d_u-1))$.  Their formula comes from a fit to simulation results \cite{kansal_computer_2002}. This expression is reasonable when the two particle diameters are similar in value, as they will be throughout this paper.

With everything now in place, we can write out the full expression for the Gibbs free energy per particle of the liquid state:

\begin{gather}
    g_l=k_BT \ln\rho/e-a(x)\rho+P/\rho-Ts_{mix}+\Delta f^{nb}+\Delta f^{b}\\
    a(x) = a_{uu}(1-x)^2+a_{ub}2x(1-x)+a_{bb}x^2\\
    s_{mix}=-k_B\bigg(x\ln x+(1-x)\ln (1-x)\bigg)\\
    \Delta f^{b}=2x\cdot(-\epsilon-Ts_b)\\
    s_b = \log\left[\left(\frac{2\pi}{\beta \kappa_b}\right)^{1/2}\left(\frac{\rho}{e}\right)^{1/3} \right]\\
    \Delta f^{nb} = \int_0^{\rho}(Z_{nb}-1)\frac{d\rho'}{\rho'}
\end{gather}

\subsection{Equilibrium liquid properties of the models}
While simple in concept, the model we have constructed has many material specific parameters ($d_b$, $\kappa_b$, $\epsilon_b$, $\epsilon_{uu}$, $\epsilon_{ub}$, and $\epsilon_{bb}$), which are required to describe a particular liquid material. Once the material parameters are specified, one may find stable states by minimizing the free energy per particle $g_l(\rho, x)$ with respect to the density $\rho$ and the fraction of bonding particles $x$ for any particular pressure $P$ and temperature $T$. 

For certain values of $T$ and $P$, there exist multiple (local) free energy minima as a function of $\rho$ and $x$. These correspond to the multiple equilibrium phases which are (meta)stable at that temperature and pressure. The minimum having the lowest Gibbs free energy is the equilibrium phase, while any other minima are metastable with respect to the true equilibrium phase. Of course, both liquid phases may also be metastable to the crystal, as is the case for water at low temperatures.

At values of $T$ and $P$ where the two phases have the same free energy per particle, those two phases are said to coexist. Such coexistence points include, for instance, the familiar boiling points. In fact, there is a coexistence curve of points, $P_{coex}(T)$, separating regions of the gas, low density liquid, and high density liquid in a $P$ - $T$ phase diagram. If the pressure is increased from the coexistence curve, the phase with lower density becomes metastable. Eventually, at some pressure $P_{spin}^{+}(T)$, we find a so called ``spinodal point" where $\frac{\partial P}{\partial \rho}\big|_T = 0$. Above this density, the corresponding metastable minimum no longer exists. Similarly, there is a lower curve of spinodal points $P_{spin}^{-}(T)$ where the high density phase loses stability.

\subsection{Random First-Order Transition Theory of the glass transition}\label{ssec:RFOT}
In the RFOT theory of glassy liquids \cite{kirkpatrick_scaling_1989}, we think of a liquid as being described in terms of a “global library” \cite{lubchenko_theory_2004} of glassy states. Each glassy state is a (local) minimum of the overall free energy landscape, where each particle is well localized near some location, $\bm{R}_i$ and vibrates around this location with some vibrational amplitude $\alpha_i^{-1/2}$. Below the crossover temperature to activated dynamics $T_C$, a glassy state is stable to small shifts in the particle positions, reflecting a finite free energy barrier to changing the particle positions in the frozen minimum by more than a vibrational amplitude. In the infinite dimensional mean field theory based on replica symmetry breaking \cite{kurchan_exact_2012}, the free energy barrier to escape such a state diverges, allowing the metastable glassy states in infinite dimension to be defined in a strict thermodynamic sense. In 3 dimensions, the glassy states can only be unambiguously defined on timescales shorter than the relaxation time corresponding to transitions between these states. On longer timescales, particles will have shifted by distances larger than the Lindemann length $0.1 d$ through cooperative motions, after which the liquid will now be locally well described by a different global glassy state.

The ``global library" construction \cite{lubchenko_theory_2004, bouchaud_adam-gibbs-kirkpatrick-thirumalai-wolynes_2004} is used to describe the dynamics by which a liquid or glass moves from one glassy state to another. The global library includes each glassy state, as well as the relevant thermodynamic properties of that particular state, such as its energy and vibrational entropy. The equilibrium liquid can then be thought of as a thermally averaged ensemble of these states, with a mean energy, mean vibrational entropy, and configurational entropy $S_c = N s_c$ arising from the number of glassy states $\Omega = \exp(N s_c/k_B)$ that are sampled by the liquid.

\begin{figure*}
    \centering
    \includegraphics[width=\textwidth]{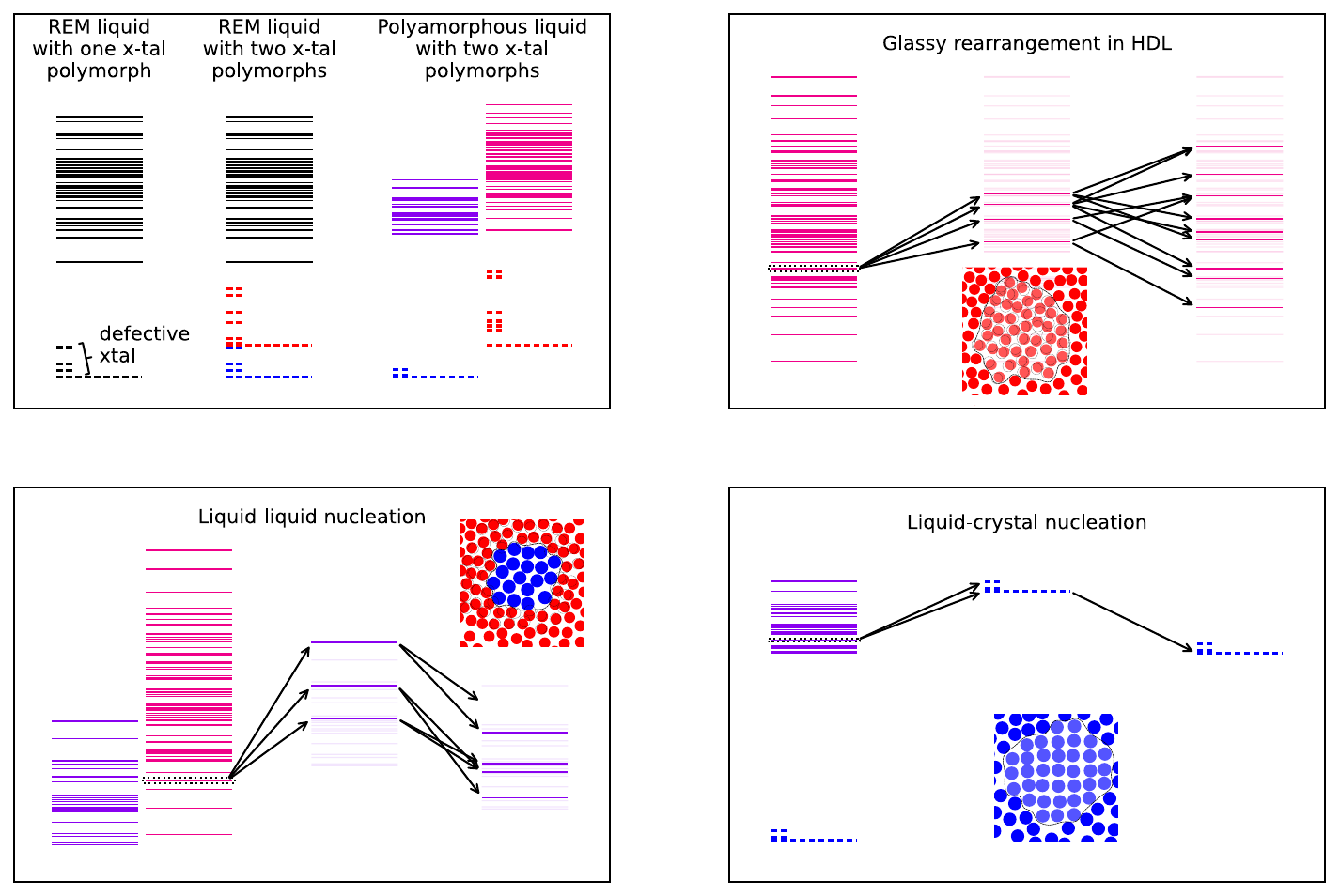}
    \caption{This schematic summarizes the library construction of the RFOT theory as it applies to a variety of materials. In the simplest case (left side of (a)), a typical liquid has a free energy landscape with roughly Gaussian energy levels, and a crystal structure at lower energies. The crystal structure also often corresponds to some deformed or defective states with imperfect crystal lattices. More commonly, (middle of (a)) multiple crystal lattices may be considered and compete with each other at different temperatures and pressures, leading to polymorphism and a multitude of crystal states. Depending on the interactions of distinct local structural motifs, there may be a distinctly bimodal, non-Gaussian free energy landscape in the liquid state(s) leading to polyamorphism (right side of (a)). Local libraries may also be constructed in order to analyze glass rearrangements in a liquid (b), liquid-liquid nucleation (c), and liquid-crystal nucleation (d). }
    \label{fig:library}
\end{figure*}

For a rearrangement starting from one particular state with a free energy $\phi_{in}^\mathrm{lib}$, we construct a series of ``local libraries" which encode all of the (entropically many) ways that a local rearrangement could occur that leave most of the glass only elastically disturbed. The local library for rearranging $N$ particles at a particular region in space is constructed by freezing in all the particles around that region, allowing some small local elastic distortion of the exterior, perhaps, except for a droplet of $N$ particles, each of which is allowed to move distances larger than a Lindemann length, as shown in figure \ref{fig:library}. The $N$ droplet particles are allowed to rearrange, finding several new local free energy minima. In general, there are $\exp(Ns_c/k_B)$ states on average in the local library of size $N$, reflecting the entropic driving force for rearrangement. The free energy of the $j$th state in the local library, $\phi_{j}^\mathrm{lib}$ has energetic and vibrational entropic components which can be related to the intermolecular forces. When a library state is locally inserted into the frozen initial state, the free energy cost is $\phi_{j}^\mathrm{lib}-\phi_{in}^\mathrm{lib}=\Phi_{j}^\mathrm{bulk}-\Phi_{in}^\mathrm{bulk}+\Gamma_{j,in}$ where $\Gamma_{j, in}$ is a surface term arising from the mismatch between the two structures at their boundary. This library construction due to Lubchenko and Wolynes \cite{lubchenko_aging_2018} resembles the construction invoked by Biroli and Bouchaud to quantify a point-to-set correlation length \cite{biroli_thermodynamic_2008}.

The ``mismatch energy'' is ultimately the energetic component opposing glassy rearrangements. In dynamics the mismatch energy functions in a way analogous to the way the bulk surface tension opposes nucleation in ordinary classical nucleation theory as described by Turnbull and Fisher \cite{turnbull_rate_1949}. Unlike ordinary surface energies, which scale as the surface area of a nucleating cluster $\gamma N^{2/3}$, the mismatch energy in the RFOT theory (near the ideal glass transition \cite{kirkpatrick_scaling_1989}) instead scales as $\gamma N^{1/2}$. This scaling is consistent with the presence of an underlying ideal glass transition at $T_K$ described by hyperscaling and a discontinuous heat capacity jump. This different slower scaling with $N$ arises because near such an ideal glass transition, there is a multiplicity of accessible states at the droplet interface. Even if the particular chosen droplet state were particularly energetically incompatible with the frozen state outside the droplet, there is some third state which can be interpolated between the two states with a lower energy cost. The arguments of Villain \cite{villain_equilibrium_1985} for the random field Ising model, when applied to a structural glass \cite{kirkpatrick_scaling_1989}, also suggest that the renormalized surface tension should scale as $\gamma N^{1/2}$ just as comes from the scaling argument. Calculations of the point-to-set length by Cavagna, Grigera, and Verrocchio \cite{cavagna_mosaic_2007} show a mismatch energy scaling in a way consistent either with $N^{2/3}$ or $N^{1/2}$.

The prefactor $\gamma$ can be estimated by relating the mismatch energy at short length scales to the entropic cost of localizing a particle, as was shown by Xia and Wolynes \cite{xia_fragilities_2000} using a density functional framework. Near $T_K$, other contributions to the interface mismatch free energy cancel out or vanish, resulting in an approximately universal estimate of $\gamma_{XW} = \gamma_0 k_B T =\frac{3\sqrt{3\pi}}{2}k_BT\ln\left(\frac{(a/d_L)^2}{\pi e}\right) $, where $d_L/a\approx 0.1$ is the familiar Lindemann ratio. The Lindemann ratio varies weakly with temperature and with material properties, but since $\gamma_{XW}$ depends only logarithmically on the Lindemann ratio, we can reasonably neglect this variation. The resulting ``chemical universality'' allows the RFOT theory to make predictions across many materials with different intermolecular interactions that compare quite well to experiment \cite{lubchenko_theory_2007}. Rabochiy and Lubchenko have also shown how to obtain $\gamma$ using elastic and structural data \cite{rabochiy_liquid_2012, rabochiy_microscopic_2013}. The universality of the Xia-Wolynes result makes it easier to employ in the present paper, which we do therefore for concreteness.

Now that we have obtained the free energetic difference between the initial glassy states and individually glassy states in the local library, we can construct the free energy for rearranging a droplet of $N$ particles from the initial glassy state. Since there are $\Omega=\exp(Ns_c/k_B)$ glassy states accessible for the droplet of size $N$, motions in the global library are driven by a configurational entropy and takes the form $F(N) = -T s_c N+\phi_{j}^\mathrm{lib}-\phi_{in}^\mathrm{lib} = N f_{liquid}(T, P) - N f_{glass}(T_{in}, P_{in}) + \gamma_{XW} N^{1/2}$. $f_{glass}$ is the free energy per particle of an individual glassy state. The typical value for $f_{glass}$ can be computed from the microscopic forces using the self-consistent phonon theory \cite{stoessel_linear_1984}, as we describe below. $f_{liquid} = f_{glass}-T s_c$ is the liquid free energy per particle, described in section \ref{ssec: model}.

For glassy rearrangements in an equilibrated liquid, $T_{in} = T$ and $P_{in} = P$ and the rearrangement free energy reduces to $F(N) = -T s_c N+\gamma_{XW} N^{1/2}$. The free energy barrier, calculated from $dF(N^\ddag)/dN = 0, F^\ddag = F(N^\ddag)$ results in an Adam-Gibbs-like viscosity $\tau = \tau_0 \exp(32 k_B/s_c)$, which when supplied with the empirical relation $s_c = \Delta c_p (T_g) \frac{T_g}{T_K}\bigg(1-\frac{T_K}{T}\bigg)$ results in the Vogel-Fulcher-Tamman law $\tau = \tau_0 \exp\bigg(\frac{D T_K}{T-T_K}\bigg)$. From these relations, the RFOT theory establishes a quantitative and predictive connection between many thermodynamic and kinetic measurements \cite{xia_fragilities_2000, xia_microscopic_2001, lubchenko_theory_2004, stevenson_thermodynamickinetic_2005}.

The above construction implicitly considers rearrangements as occurring within a roughly spherical, compact cluster. Rearrangements may also occur with varying shapes, and in particular rearrangements may be ramified and string-like. Such rearranging clusters were considered by Stevenson, Schmalian, and Wolynes \cite{stevenson_shapes_2006}, who showed that while such structures accrue a higher energetic penalty due to the increased surface area of the non-spherical clusters, they have an additional entropic driving force for rearrangement due to the many configurational arrangements of a string-like cluster. Above a particular crossover temperature $T_C$, the rearrangement free energy for string-like rearrangements is barrierless and downhill. Thus $T_C$ marks the crossover between collisional dynamics at high temperature and activated dynamics at low temperature. This argument suggests the crossover occurs at a nearly universal configurational entropy $s_c=1.28 k_B$ per bead where relaxations are in the microsecond range, consistent with observations of the timescale for the dynamic crossover \cite{novikov_universality_2003}. String-like rearrangements are also responsibly for secondary relaxations \cite{stevenson_universal_2010} below $T_C$.

\subsection{Self-Consistent Phonon Theory}
The free energy of an individual glassy state can be calculated using a simple approximation known as the self-consistent phonon method. The method originally dates to Fixman's theory of freezing \cite{fixman_highly_1969}, and has since been further developed and applied to the hard sphere glass \cite{stoessel_linear_1984}, a network glass with hard sphere repulsions \cite{hall_microscopic_2003}, a Lennard-Jones glass \cite{hall_intermolecular_2008}, motorized biological assemblies \cite{shen_statistical_2006, wang_interplay_2011, wang_tensegrity_2012}, and most recently to mixtures of network-forming and non-network-forming particles \cite{brown_bonding_2025}. Self-consistent phonon theory operates by assuming a particular fixed arrangement of particles, which are then allowed to vibrate around their given coordinates $\bm R_i$, providing a Gaussian density profile $\rho_i(\bm r)=(\alpha_i/\pi)^{3/2}\exp(-\alpha_i(\bm r-\bm R_i)^2)$. Each particle experiences a local effective potential $\exp(\beta V_i^\mathrm{eff})=\prod_{j\neq i}\int d\bm{r_j} \rho_j(\bm{r_j}) \exp\left(-\frac{\beta}{2} V_{ij}(
\bm{r}_j-\bm{R}_i)\right)$ where $V_{ij}$ is the interparticle interaction, which is averaged over the density fluctuations of nearby particles. Using the mean-field pair correlation function, the typical particle vibrates as $
\alpha=\frac{\rho}{6}\int 4\pi R^2dR g(R) \nabla^2 V^\mathrm{eff}(R)$ and has a mean free energy

\begin{eqnarray}
    f_g=&k_B T\rho\int d\bm{R} g(R;\rho) \sum_n p_n \beta V^\mathrm{eff}(\bm{R};\alpha_n)\nonumber\\
    &-k_BT \sum_n p_n \frac{3}{2}\ln \frac{\pi}{\alpha_n}-Ts_\mathrm{mix}
\end{eqnarray}

where a particle with $n$ bonds has a vibration tensor $\alpha_n$. Note that in our present model, the mixing entropy may be accessed even within a glassy state because we allow the bonding and non-bonding particles to interconvert freely. This would be different for a permanently bonded ``vulcanized'' glass.

The radial distribution function $g(R)$ can be computed using methods similar to those used in section \ref{ssec: model} for computing the thermodynamics of the liquid state. Neglecting specific bonding constraints, the structure of the liquid should resemble that of a binary hard sphere mixture with two particle radii. We use the radial distribution function of Mansoori, Carnahan, Starling, and Leland \cite{mansoori_equilibrium_1971} as modified by Grundke and Henderson \cite{grundke_distribution_1972}. These functions are completely analytical and can be calculated with computer code which has been helpfully made freely available \cite{smith_fortran_2008}. The mixture formulas are analogous to the single-particle hard sphere distribution functions due to Henderson and Grundke \cite{henderson_direct_1975} and Verlet and Weis \cite{verlet_equilibrium_1972}. The effective hard sphere fluid has a temperature dependent diameter, as shown in equation \ref{eq: Barker-Henderson}.

The self consistent phonon method can also be used to study crystal phases directly, by using the crystal structure to construct a pair correlation $g(\bm{R})$ as a sum of Dirac delta functions. We discuss this point further in the supplemental information where we describe crystal melting transitions, since polyamorphism often occurs in a ``no-mans' land'' where crystallization is hard to avoid in the laboratory.

\subsection{An aside on units}
Throughout this paper, when discussing results, we have in each case assigned enough dimensionful microscopic parameters to fully constrain the model. For that reason, we therefore report results using the units such that $d_u = 1$ and $\epsilon_{uu} = 1$ for all reported results. In particular, the pressure and temperature are reported in these microscopic units. When comparing with actual molecular materials we should then bear in mind that a temperature of $T \approx \epsilon_{uu}/k_B$ is hot enough so that thermal fluctuations may break a Lennard Jones interaction. An external pressure of $\epsilon_{uu}/d_u^3$ corresponds to the compressive microscopic forces holding such a pure Lennard Jones fluid together. We caution that in real materials, the microscopic length and energy scales vary widely. Notably so when comparing to the behavior of water and silicates. Rather than thinking directly of the length and energy scales, when discussing the interplay of polyamorphism and the glass transition, we emphasize that it is important to compare the relative energy scale of bonding and nonbonding forces, or the relative densities of bonded and nonbonded structures.

\section{Points of interest on the phase diagram}\label{sec: points of interest}
For the sake of concreteness, throughout this section, we will employ the parameters $d_b/d_u = 1.2$, $\epsilon_{bb}/\epsilon_{uu} = 1$, $\epsilon_{ub} = 0$ and $\epsilon_b/\epsilon_{uu} =7$ unless otherwise noted. These parameters result in a typical phase diagram having 3 amorphous fluid phases: a gas and two distinct liquids. 
\subsection{Coexistence curves, spinodals, and critical points}
\begin{figure*}
    \includegraphics[width=\linewidth]{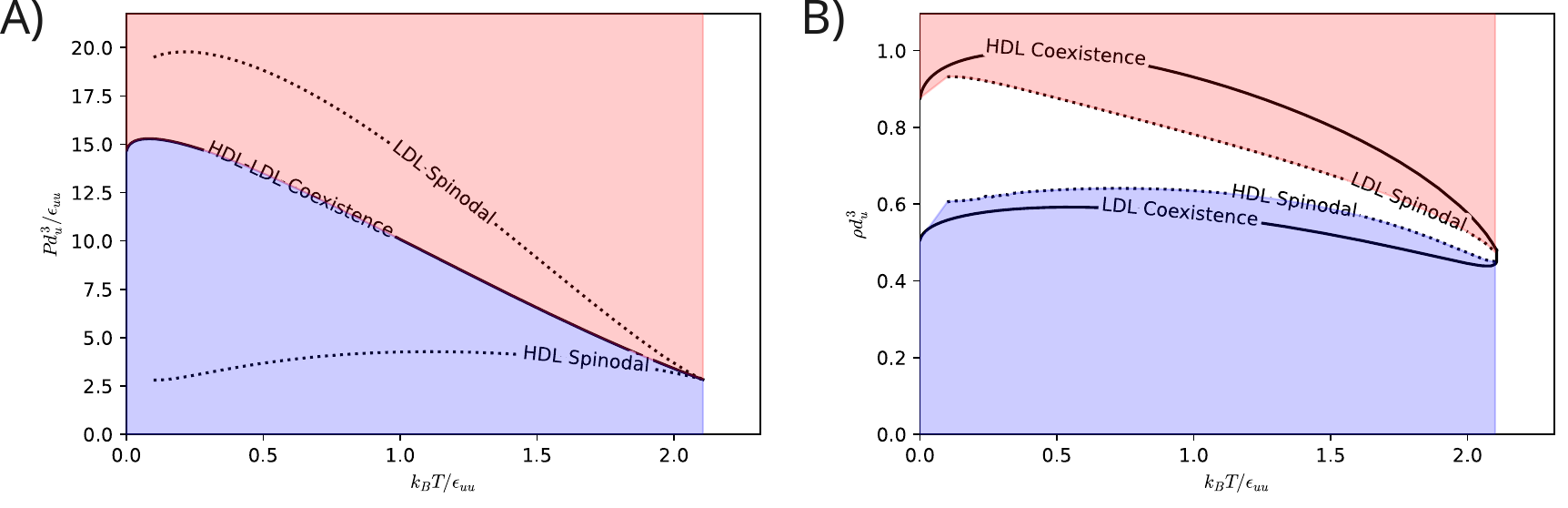}
    
    \caption{Spinodal points and coexistence curves for $\epsilon_{ub} = 0, \epsilon_{bb}/\epsilon_{uu} = 1, \epsilon_b/\epsilon_{uu} = 7$. These parameters result in a material with $T_c^{LL}$ well above $T_g$, so that critical and glassy properties can be well-separated. The coexistence curve (shown as a solid line) marks the pressure (a) at which the LDL and HDL may coexist stably, i.e. they have the same chemical potential. Similarly, (b) the densities of the two coexisting phases are shown. At higher pressures, the LDL becomes metastable. At sufficiently high pressure, the LDL becomes completely unstable at the spinodal point, marked with a dotted line. Similarly, at lower pressures the HDL becomes metastable and then unstable at spinodal marked with another dotted line.}\label{fig: base spinodal}
\end{figure*}
To confirm the presence of a liquid-liquid phase transition in the model, we first locate the coexistence and spinodal curves, shown in figure \ref{fig: base spinodal}. Below $T = T_c^{LL} \approx 2.1 \epsilon_{uu}$, two distinct liquid phases can be manifested. The pressure required for these phases to coexist increases as the temperature is decreased. This indicates, as is to be expected, that the low density liquid is a low energy, low entropy phase relative to the high density liquid. As is true in typical network liquids, when bonds are formed in the low density liquid, both the energy and entropy per particle are decreased. The coexistence densities of the two liquid phases become equal at the critical point. Except just near the critical point, both densities increase as the temperature is further decreased. At low temperatures $T\lesssim0.1 \epsilon_{uu}$, the predicted coexistence pressure exhibits a turnover. This is primarily because the Barker-Henderson effective hard sphere diameter begins to saturate at the minimum of the Lennard-Jones potential. Data below this temperature are not accurate due to this breakdown in the approximations we have used. In earlier work, Hall and Wolynes \cite{hall_intermolecular_2008} used the Kang, Ree, Ree separation of the potential \cite{kang_perturbation_1986}, which does not have this deficit. Fortunately, as we will see shortly, this turnover also occurs below the predicted Kauzmann temperature $T_K$, so the ``equilibrium" liquid values computed from this free energy would not be meaningful in any case.

As $T\rightarrow T_c^{LL}$, the coexistence pressure and both spinodal pressures converge onto a single critical pressure. Similarly, the coexistence and spinodal densities converge onto a single critical density. This critical point, analogous to the traditional liquid-gas critical point, marks the maximum temperature at which two distinct dense liquid phases can be simultaneously present at equilibrium. Near this critical point, we expect there to be long range critical fluctuations resulting in diverging length and time scales. We note however, that these divergences are distinct from the divergences present in the Random First-Order Transition theory as the liquids approach the Kauzmann points. Constructing a unified theory of the two phase critical fluctuations in the vicinity of a glass transition is thus an interesting problem which we will not discuss in depth in the present work.

\subsection{Glass transition temperature}
\begin{figure*}
    \centering
    \includegraphics[width=\linewidth]{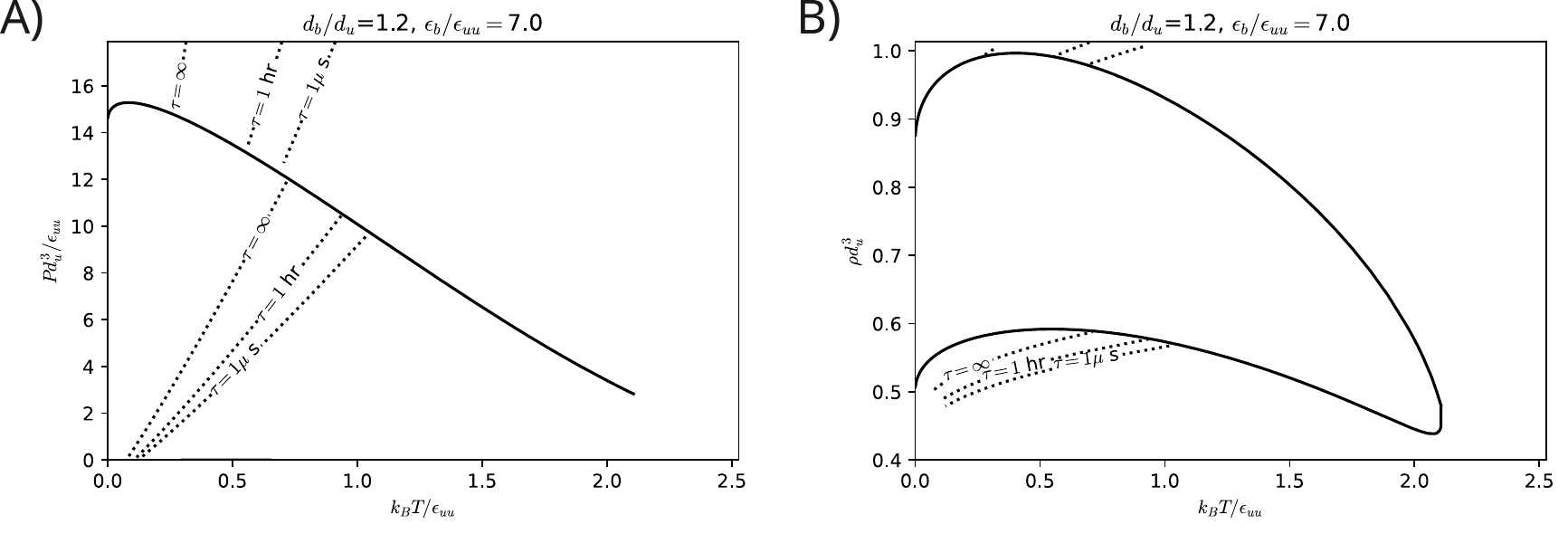}
    
    \caption{Here, we plot lines of constant configurational entropy, $s_c$ vs (a) pressure and temperature, or (b) density and temperature. We plot curves at $s_c$ values corresponding to rearrangement timescales of 1 $\mu$s, 1 hr, and $\infty$. These timescales define the characteristic temperatures for the string crossover $T_C$, the laboratory glass transition $T_g$, and the Kauzmann transition $T_K$. Each temperature is pressure dependent, as dictated by the equation of state.}\label{fig: Tc, Tg, Tk}
\end{figure*}

The temperature at which a liquid falls out of equilibrium and forms a glass depends on a combination of factors, including the exact protocols by which pressure and temperature changes are applied to the liquid, and over what timescale. In addition, the material properties depend on the microscopic inputs $\epsilon_b$, $\epsilon_{XY}$, and $d_b$ in our model. The simplest heuristic is that whenever the timescale of rearrangements in the liquid is longer than the timescale over which the applied temperature or pressure is changed, then the liquid falls out of equilibrium. This fact is reflected in the various definitions of the laboratory $T_g$: some definitions use an hour timescale, some a 100 s timescale, or many others, depending on timescale of the experiment itself. 

Within the RFOT theory, the rearrangement timescale primarily depends on the configurational entropy per particle since the mismatch parameter $\gamma$ is nearly universal. In figure \ref{fig: Tc, Tg, Tk}, we thus show curves of constant rearrangement timescale, or equivalently constant configurational entropy. The curves labeled $\tau  = 1$ $\mu$s mark the so called string crossover of Stevenson, Schmalian, and Wolynes \cite{stevenson_shapes_2006} $s_c = 1.28 k_b$ per bead. At this crossover, the rearrangement timescales are approximately 1 $\mu$s as is confirmed empirically \cite{novikov_universality_2003}. 1 $\mu$s is also a typical timescale accessed in modern computer simulations. We similarly mark the curve of $T_g$ at the 1 hr timescale, a value commonly used in laboratory studies, and the curve of $T_K$ where the configurational entropy vanishes so the rearrangement timescales would diverge to infinity. The correspondence between timescales and configurational entropy follow from the analysis in section \ref{ssec:RFOT}, where $\tau_\alpha/\tau_0 = \exp(32 k_B/s_c)$, if $s_c$ is measured per ``bead" (i.e. the effective sphere-like rearranging units of a molecule)\cite{stevenson_thermodynamickinetic_2005}.

The precise values of $s_c$ depend in a complex way on $x, \rho$, and $T$, and indirectly on $P$, which itself determines the density. Nevertheless, we note based on figure \ref{fig: Tc, Tg, Tk} (b) that the configurational entropy is most strongly density dependent. The curves of constant configurational entropy are nearly horizontal on the $\rho-T$ plane, indicating that small changes in density result in large changes in configurational entropy, while small changes in temperature have a smaller effect.

\subsection{Water-like anomalies}\label{ssec: water-like anomalies}
\begin{figure*}
    \centering
    \includegraphics[width=\linewidth]{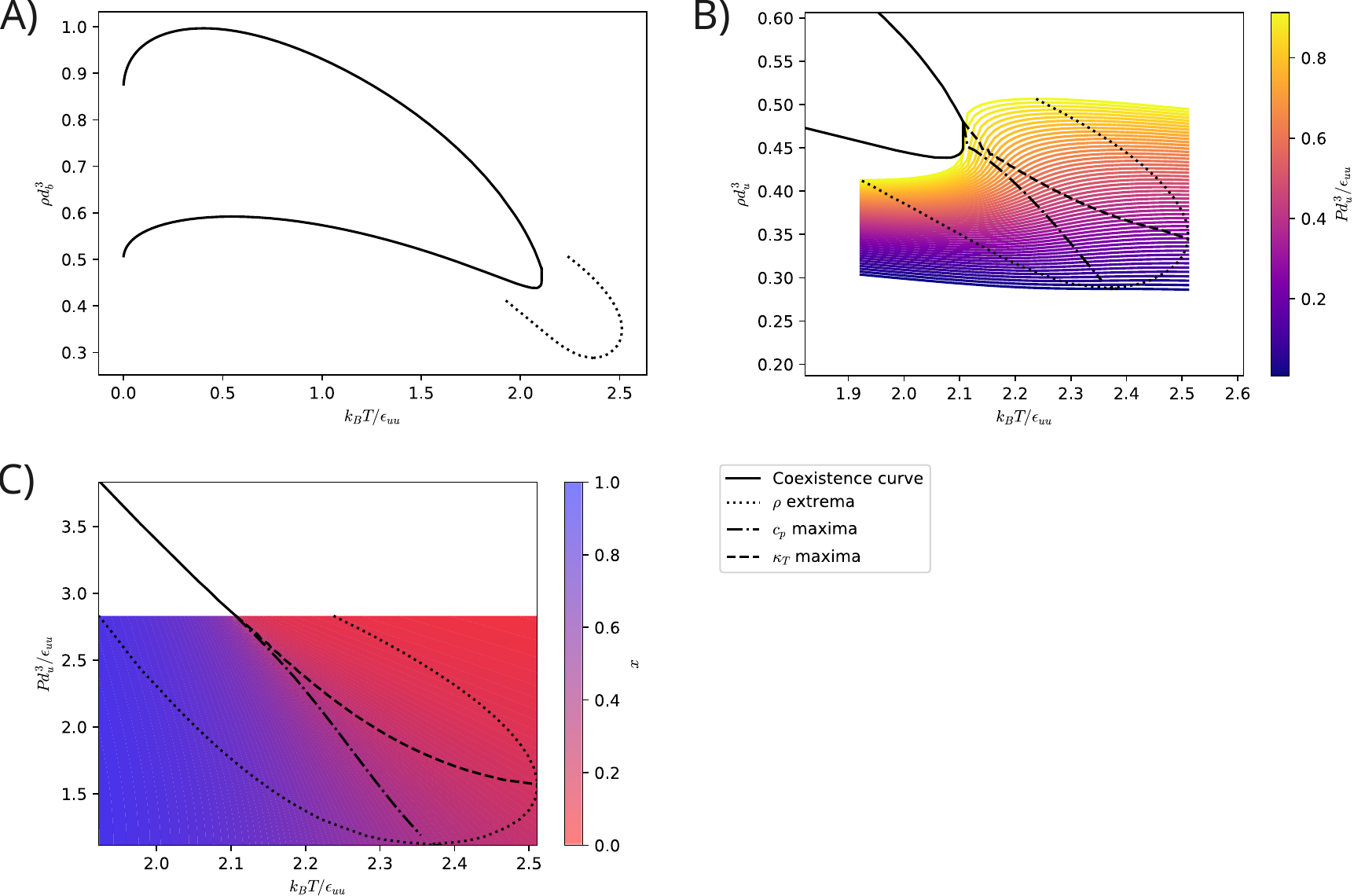}
    \caption{Near the critical point, the model features water-like anomalies. (a-b) The density shows a maximum and minimum with respect to temperature (c) other thermodynamic quantities, such as the heat capacity $c_P$ and isothermal compressibility $\kappa_T$ show a curve of maxima. The thermodynamic anomalies occur in the region of the phase diagram where $x$ varies most strongly with either $T$ or $P$, as shown by the color in (c).}\label{fig: density extrema}
\end{figure*}

The thermally driven bond breaking present within our model allows it to capture the behavior of water-like anomalies found in common network materials. There are several anomalies in different thermodynamic quantities which arise from a negative thermal expansion coefficient. In most liquids, the thermal expansion coefficient $\alpha = \rho\frac{\partial (1/\rho)}{\partial T}\big|_P$ takes a positive value, and the material increases in volume when heated. In network glasses under certain pressure and temperature conditions, heating the material may instead allow bonds to break, thus allowing particles to pack together more tightly, thus reducing the volume. Negative thermal expansion occurs in the region of the phase diagram where $x$ is varying most quickly with respect to variations in temperature and pressure. Notably, above the liquid-liquid critical temperature $T_c^{LL}$, $x$ must vary continuously, but below $T_c^{LL}$, $x$ jumps discontinuously with pressure at the coexistence curve. As such, near the critical point we have a quickly varying $x$ and thus negative thermal expansion and other anomalies.

The first, and most well known thermodynamic anomaly present in network liquids is a density maximum as a function of temperature. At atmospheric pressure, water is famously most dense at 4$^\circ$C, and expands upon both cooling and heating from that point. In fact, there is a curve of temperatures of maximum density for a range of pressures, as shown in the right-hand branch of the $\rho$ extrema curve of figure \ref{fig: density extrema}. In addition, upon further cooling from the density maximum, the liquid first expands until most of the bonds are formed and the density reaches a minimum value. Further cooling below the density minimum results in contraction. Negative thermal expansion occurs precisely between the density minima and maxima, where bonds break most quickly with heating.

In addition to these density extrema, the liquid liquid phase transition results in a line of maxima of the isothermal compressibility $\kappa_T = -\rho\frac{\partial (1/\rho)}{\partial P}\big|_T$. The compressibility is maximized along isobars, resulting in the curve shown in figure \ref{fig: density extrema}. For glass physics, we note that the heat capacity $ c_P = -T\frac{\partial^2 g_l}{\partial T^2}\big|_P$ has a maximum when computed along an isotherm, as shown in figure \ref{fig: density extrema}. This would indicate a varying apparent fragility within the RFOT theory, as the fragility is related to $\Delta c_P$, the part of the heat capacity associated with the configuarational entropy. Similar anomalies in the density, heat capacity, and compressibility were attributed to such an underlying liquid-liquid phase transition in water by Poole et al. \cite{poole_phase_1992}. 

\subsection{Fragility}
Supercooled liquids and glasses have often been characterized based on the temperature dependence of the viscosity into ``strong" and ``fragile" glass formers \cite{angell_formation_1995}. Strong liquids (including silica and water at atmospheric pressure) feature a nearly Arrhenius viscosity and a very low drop in heat capacity per particle $\Delta c_p$ upon crossing the glass transition. $\Delta c_p$ is also called the ``configurational" heat capacity, corresponding to the configurational entropy, and is the difference between the full heat capacity of the liquid and the vibrational part of the heat capacity.

The fragility can be quantitatively characterized by the $m$ value 
\begin{equation}
    m = \frac{d(\log_{10}\tau_{\alpha}/\tau_0)}{d(T_g/T)}
\end{equation}
which measures the slope of the viscosity on an Arrhenius plot, where log viscosity is plotted against $1/T$. and is thus related to the aparent activation enthalpy at the glass transition temperature. Glass formers display a broad spectrum of $m$ values \cite{stevenson_thermodynamickinetic_2005}, and glass formers should not be classified in a binary way. Indeed, a ``strong'' glass former such as silica becomes quite fragile merely by the addition of a few percent of alkali oxides, as we explained in our earlier paper \cite{brown_bonding_2025}. 

Within the RFOT theory, the variation of fragility is naturally explained in terms of thermodynamic variations. As discussed in section \ref{ssec:RFOT}, the timescale for glassy rearrangements is roughly $\tau_\alpha=\tau_0\exp(32k_B/s_c)$. Noting that the heat capacity drop at the glass transition is given by $\Delta c_p(T_g) = T_g\frac{ds_c(T_g)}{dT}$, we find a linear relationship between the fragility and the heat capacity drop $m\approx 22 \Delta c_p/k_B$ \cite{stevenson_thermodynamickinetic_2005}, where $\Delta c_p$ should be measured per ``bead". This simple relation is a result of the approximate universality of $\gamma$ suggested by Xia and Wolynes \cite{xia_fragilities_2000}.

We can compute the configurational heat capacities of the liquids in our model quite easily by noting how the configurational entropy varies with temperature, as computed in section \ref{ssec:RFOT}. We noted also in section \ref{ssec: water-like anomalies} that the total heat capacity diverges at $T_c^{LL}$ and passes through a peak value as a function of pressure. It is thus natural to also look for a peak in the configurational heat capacity.

\begin{figure*}
    \centering
    \includegraphics[width=\linewidth]{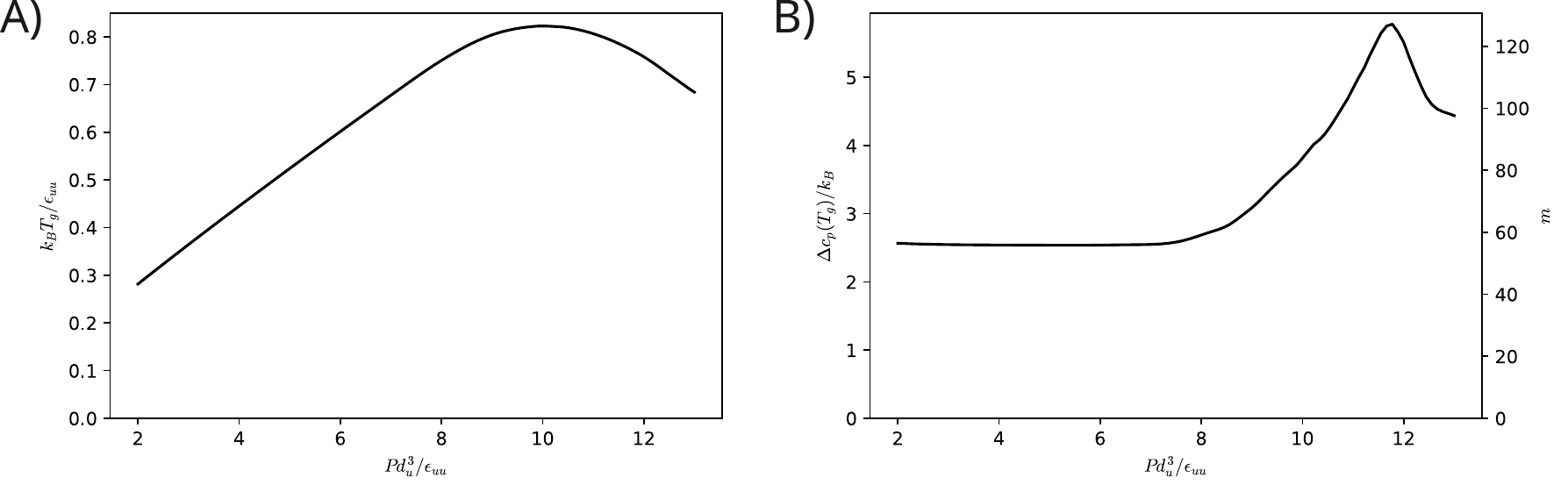}

    \caption{Here, we plot (a) the glass transition temperature $T_g$ and (b) the heat capacity drop at $T_g$ as a function of pressure. For this case, we use $\epsilon_{ub}/\epsilon_{uu} = 0.9$, which results in a $T_g$ just above $T_c^{LL}$. We can see a broad maximum in the glass transition temperature and a narrower peak in the configurational heat capacity, both signatures of the underlying liquid-liquid phase transition.}\label{fig: dcp maximum}
\end{figure*}
We explore now how different phase diagrams yield distinct glassy behavior. In figure \ref{fig: dcp maximum}, we plot the glass transition temperature and the configurational heat capacity at $T_g$ as a function of pressure. For these plots, we use a value of $\epsilon_{ub}/\epsilon_{uu} = 0.9$ rather than $\epsilon_{ub} = 0$, which results in $T_g$ values which are above the liquid-liquid critical point and in the region of water-like anomalies. Both the glass transition temperature and the fragility display non-monotonic behavior resulting from the water-like thermodynamic anomalies present near $T_g$. In particular, the theory predicts for this parameter choice a broad peak in $T_g$ as a function of pressure, with the maximum value of $T_g$ being almost 3 times larger than the lower $T_g$ values. This is a much wider range of glass transition temperature variation than is typically seen in real materials. The configurational heat capacity has a sharper peak at around $P d_u^3/\epsilon_{uu} = 11$, where $T_g$ is near the liquid-liquid critical point. The high configurational heat capacity corresponds to quickly varying thermodynamic variables, like $\rho$ and $x$, which change discontinuously quickly across the phase transition, and still rather just above $T_c^{LL}$ At lower pressures, $Pd_u^3/\epsilon_{uu}\lesssim 8 $, below the peak of $\Delta c_p$, the liquid obtains a lower density and bonds are allowed to form, resulting in a relatively low configurational heat capacity and a low $m$ value. Many real network liquids, including water and silica, are strong liquids at atmospheric pressure. The peak fragility of $m\approx 125$ is among the highest fragilities in real materials. For comparison, OTP, a well-studied fragile glass former, has a measured fragility of $m=81$ \cite{wang_direct_2002}. At higher pressures $Pd_u^3/\epsilon_{uu}\gtrsim 13$, above the peak $\Delta c_p$ the fragility of our network liquid model reduces somewhat, but the liquid remains quite fragile, in fact a bit more fragile than OTP for these parameters. Variations in fragility due to variations in pressure, similar to those seen here, have been termed a ``strong-to-fragile'' transition \cite{angell_water_1993, ito_thermodynamic_1999} in water, and may occur discontinuously as the liquid passes through a liquid-liquid phase transition, or smoothly as seen here.

\section{Exploration of model parameters} \label{sec: investigation of model parameters}
The model, as described thus far, has a rich set of input parameters in the form of the different energy scales $\epsilon_b$ and $\epsilon_{XY}$. The properties of a particular ``material", and thus a particular liquid-liquid phase diagram, is determined by the ratio of these energy scales. We can learn about the different possible behaviors of the model by tuning each ratio of energy scales in turn, while keeping the others fixed.

\subsection{Tuning $\epsilon_b$}
\begin{figure*}
    \centering
    \includegraphics[width=\linewidth]{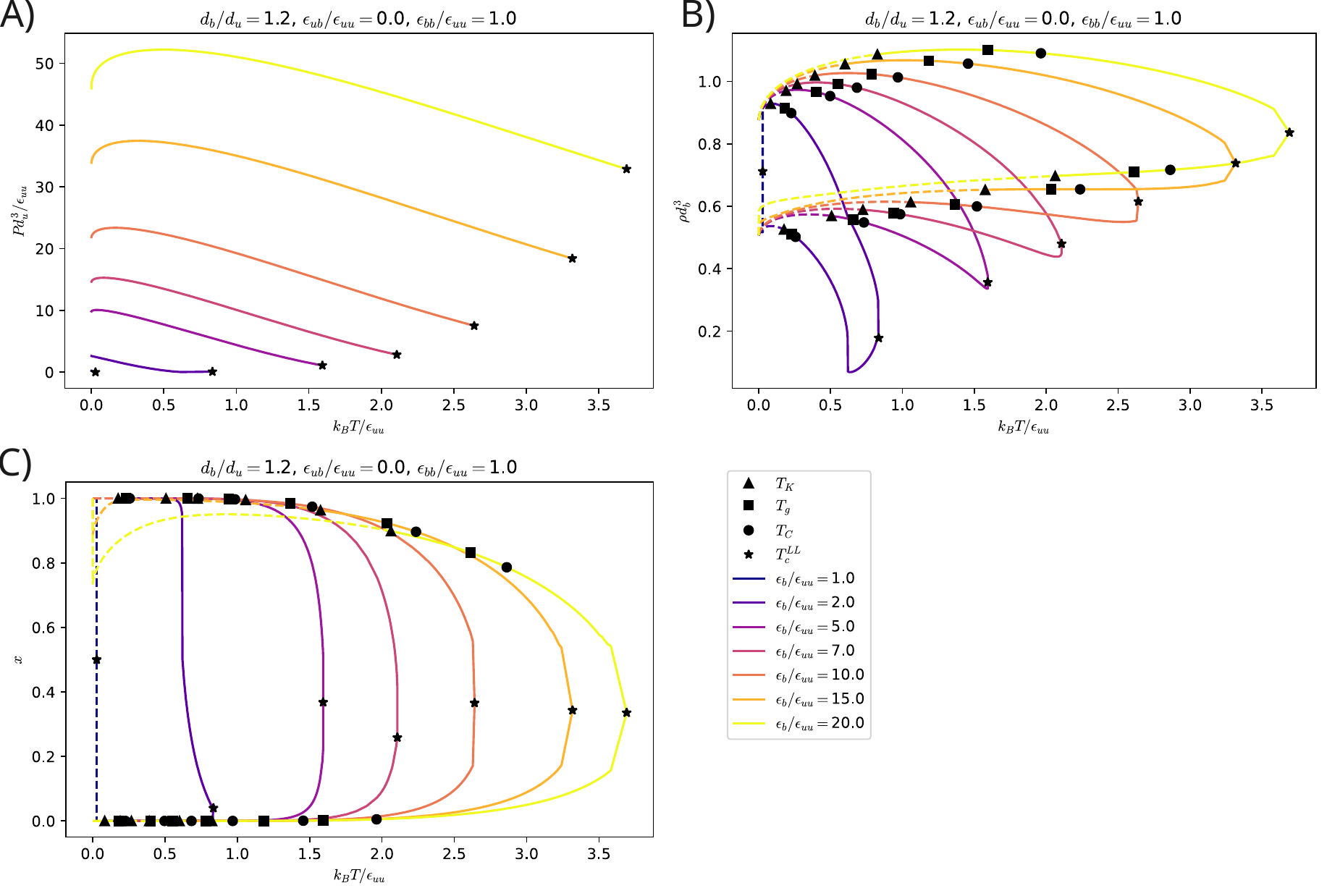}

    \caption{Here, we plot the dependence of the coexistence curves on $\epsilon_b$. In all cases, $\epsilon_{bb}/\epsilon_{uu} = 1$ and $\epsilon_{ub} = 0$. The curves for $\epsilon_{b}/\epsilon_{uu} = 7$ correspond to the material discussed in figures \ref{fig: base spinodal}, \ref{fig: density extrema}, and \ref{fig: Tc, Tg, Tk}. The coexistence curves are shown in terms of (a) pressure vs temperature, (b) density vs temperature, and (c), bonding fraction vs temperature. As $\epsilon_b$ is increased, the LDL is stabilized relative to the HDL. To account for this, the coexistence shifts to higher pressures. At sufficiently low values of $\epsilon_b$, there is no liquid-liquid phase transition, as there is no energetic advantage for the liquid to form an LDL. Three characteristic temperatures for local rearrangements are shown on the corresponding curves: the string crossover temperature, $T_C$; the laboratory glass transition temperature, $T_g$; and the Kauzmann temperature, $T_K$. Each of these temperatures shifts upward as $\epsilon_b$ increases.}\label{fig: eps series}
\end{figure*}

First, we tune the strength of the bonds, $\epsilon_b$, while keeping the Lennard Jones interactions fixed, with $\epsilon_{bb}/\epsilon_{uu} = 1$ and $\epsilon_{ub} = 0$. Microscopically, this may be thought of as tuning the relative importance of the bonding forces vs nonbonding forces in determining the local structure and dynamics of the liquid. It would be expected, for instance, that a network liquid like SiO$_2$ would have a relatively high $\epsilon_b$ compared to a network liquid like water, whose molecules interact via weaker hydrogen bonds.

In figure \ref{fig: eps series} (a), we see that the primary effect of increasing $\epsilon_b$ is to determine the pressure scale needed to access the high density liquid. The high density liquid is only stable so long as the pressure is high enough to overcome the energy of the bonds. Notably, for very low values of $\epsilon_b$, there is no liquid-liquid phase transition. This is because the low density, bonded phase is only stable relative to the high density liquid when under negative pressure, in which case it is also metastable to the gas. Additionally, as shown in figure \ref{fig: eps series} (b), materials with a higher $\epsilon_b$ have both higher coexistence pressures and higher coexistence densities. The characteristic temperatures $T_K$, $T_g$, and $T_C$ of more strongly bonded materials are also higher than those for weakly bonded materials.

\subsection{Tuning $\epsilon_{bb}/\epsilon_{uu}$}
\begin{figure*}
    \centering
    \includegraphics[width=\linewidth]{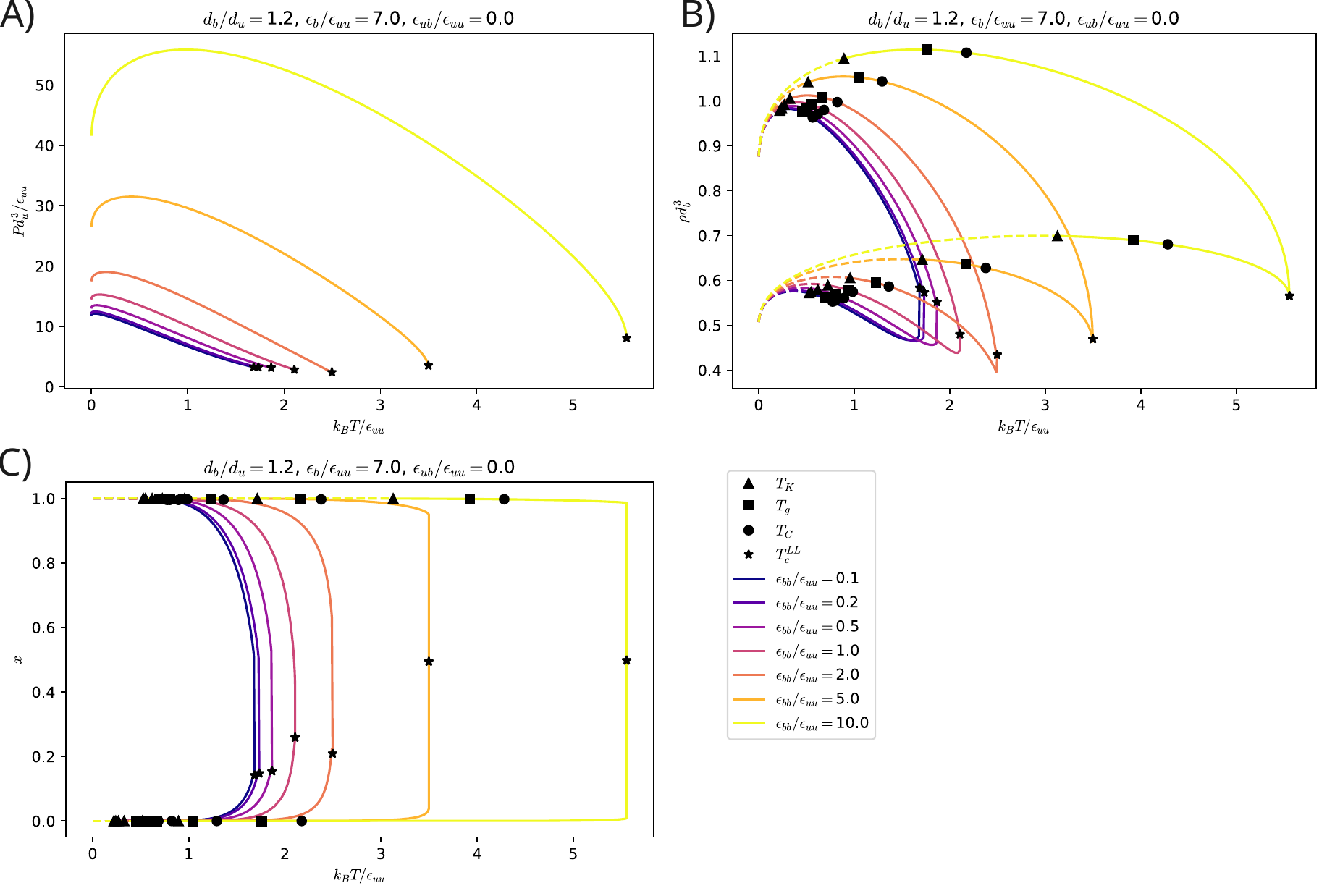}

    \caption{Here, we plot the dependence of the coexistence curves on $\epsilon_b$. In all cases, $\epsilon_{b}/\epsilon_{uu} = 7$ and $\epsilon_{ub} = 0$. The curves for $\epsilon_{uu}/\epsilon_{uu} = 1$ correspond to the material discussed in figures \ref{fig: base spinodal}, \ref{fig: density extrema}, and \ref{fig: Tc, Tg, Tk}. The coexistence curves are shown in terms of (a) pressure vs temperature, (b) density vs temperature, and (c), bonding fraction vs temperature. As $\epsilon_{11}$ is increased, the LDL is stabilized relative to the HDL. To account for this, the coexistence shifts to higher pressures. Similarly, the density of the LDL shifts higher, as the greater $\epsilon_{bb}$ value provides additional attractive forces. Three characteristic temperatures for local rearrangements are shown on the corresponding curves: the string crossover temperature, $T_C$; the laboratory glass transition temperature, $T_g$; and the Kauzmann temperature, $T_K$. Each of these temperatures shifts upward as $\epsilon_{bb}$ increases.}\label{fig: a22 series}
\end{figure*}

Tuning the van der Waals attraction of the bonding particles relative to that of the nonbonding particles has a similar effect on the phase diagram as tuning $\epsilon_b$. This is natural; both the bonding term and the van der Waals term enter into the free energy in similar ways, as long as the density varies slowly with pressure and temperature. The high density liquid is only stable at sufficiently high pressures, where the pressure is able to overcome the lower energy of the low density liquid. Materials with higher $\epsilon_{bb}/\epsilon_{uu}$ ratio thus also have higher coexistence pressures, densities, $T_K$, $T_g$, and $T_C$, as was the case for high $\epsilon_b$ materials.

\subsection{Tuning $\epsilon_{ub}/\epsilon_{uu}$}
\begin{figure*}
    \centering
    \includegraphics[width=\linewidth]{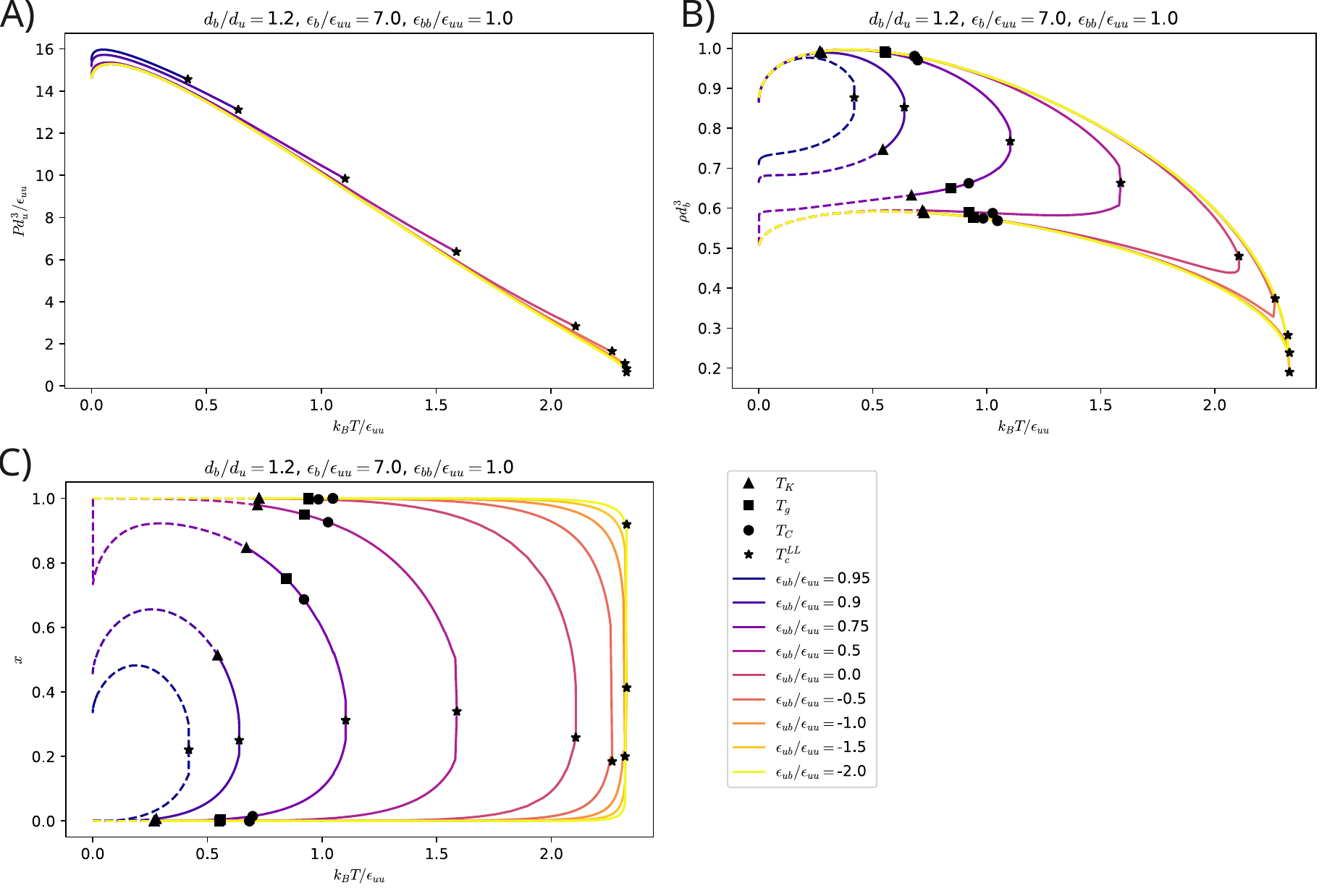}

    \caption{Here, we plot the dependence of the coexistence curves on $\epsilon_{ub}$. In all cases, $\epsilon_{bb}/\epsilon_{uu} = 1$ and $\epsilon_{b} = 7$. The curves for $\epsilon_{ub} = 0$ correspond to the material discussed in figures \ref{fig: base spinodal}, \ref{fig: density extrema}, and \ref{fig: Tc, Tg, Tk}. The coexistence curves are shown in terms of (a) pressure vs temperature, (b) density vs temperature, and (c), bonding fraction vs temperature. Large negative values of $\epsilon_{ub}$ prohibit mixing between bonding and non-bonding particles, thus extending the critical point to higher and higher temperatures. Whenever $\epsilon_{ub}\lesssim\epsilon_{uu}=\epsilon_{bb}$, mixtures of bonding and non-bonding particles are energetically unfavorable and a liquid-liquid phase separation is possible. Generally, except near the critical point, the $x$ values are quite close to 1 and 0. This separation becomes more pronounced for more highly negative $\epsilon_{ub}$. Notably, since $x\approx$ constant, the coexistence curves fall nearly along the same curve, and tuning $\epsilon_{ub}$ largely amounts to directly tuning the critical temperature. Three characteristic temperatures for local rearrangements are shown on the corresponding curves: the string crossover temperature, $T_C$; the laboratory glass transition temperature, $T_g$; and the Kauzmann temperature, $T_K$. These temperatures change only modestly as $\epsilon_{ub}$ is tuned. We see that the model permits a broad range of behaviors, where the glass transition temperature may be higher or lower than the critical temperature.}\label{fig: a12 series}
\end{figure*}

According to London's theory of dispersion forces, the attractive strength between two dissimilar particles would naturally take on a value of roughly $\epsilon_{12} = \sqrt{\epsilon_{11}\epsilon_{22}}$. This combining rule dates back to Berthelot \cite{berthelot_sur_1898} as an empirical correlation for many simple solutions. While this rule of thumb is reasonably good, of course it is not universal, especially when directional forces enter. It is instructive to put other values in by hand. The Kob-Andersen model \cite{kob_testing_1995}, a common computational model frequently used to study the glass transition, in particular describes a mixture of Lennard-Jones particles which uses an $\epsilon_{12}$ that is stronger than would be inferred from this combining rule. The additional attractive force between particles of different types in the Kob-Andersen model encourages thorough mixing of the components and destabilizes the simple crystal phases. Some more complex crystal phases \cite{middleton_crystals_2001} are thermodynamically stable to the liquid, but are instead kinetically impeded from forming due to the complexity of their crystal structures. As a consequence, using the parameters chosen in the Kob-Andersen model, simulations can be performed more easily in a uniform liquid phase without crystallization or phase separation occurring. 

In order for a liquid-liquid polyamorphic phase transition to be present within the present model, one must have $\epsilon_{ub}\lesssim \sqrt{\epsilon_{uu}\epsilon_{bb}}$. This represents an effective frustration between neighboring regions of bonded and nonbonded particles. It has been proposed by Tanaka \cite{tanaka_general_2000} that such an effective frustration is a general feature which would lead to a liquid-liquid phase transition in liquids manifesting structural motifs that have different sources of stability. Within the library picture of a liquid, a local frustration of this form leads to a splitting of the density of states. Without such a specific source of frustration, a liquid generally has a density of states with a unimodal peak, resembling a random energy model \cite{derrida_random-energy_1980}. With frustration between two different local structural motifs, the global library instead would contain two ``clumps" of states with two different mean energies, vibrational entropies, fluctuations in the energy, corresponding to the properties of those local structural motifs. This library based point of view is quite concordant with the analysis of Tanaka; indeed, the microscopic forces that result in the frustration between local structural motifs are also the microscopic origin of the splitting of the density of states into two distinct peaks. 

We note that the choices of $\epsilon_{ub}$ we explore here, which allow for the presence of a liquid-liquid phase transition, also tend to destabilize the liquid phases relative to the crystal phases, and would, in fact, on their own promote crystallization. Indeed, water is known to be a very poor glass former due to the rapidity of its crystallization upon cooling. Glass formation in water via a temperature quench has only recently been seen using extremely fast quenches in small droplets of water \cite{mayer_new_1985}, since slower temperature quenches result in crystallization. In order to study the liquid-liquid phase transition in a model like ours using a direct simulation, one likely needs to modify the model to avoid crystallization in some other way, such as by adding higher order many-body terms to destabilize crystal phases.

The results for different values of $\epsilon_{ub}$ are shown in figure \ref{fig: a12 series}. We see that tuning $\epsilon_{ub}$ has very little impact on the coexistence pressures, but instead mostly changes the temperature of the critical point. This is because (except near the critical point) the high density phase has $x\approx 0$ and the low density phase has $x\approx 1$, so the van der Waalsian interaction between the two different particles has little impact on the thermodynamics of the two phases individually. Near the critical point, we see  the system takes on intermediate values of $x$, thus a more frustrated liquid-liquid phase transition needs higher temperatures to overcome the additional energetic cost near the critical point.

While we see a broad range of $T_c^{LL}$ across different $\epsilon_{ub}$ values, there is very little variation in $T_g$. Materials with lower $\epsilon_{ub}$ have only modestly higher $T_g$. This provides us with a convenient way to tune between materials with different ratios of $T_c^{LL}/T_g$. We will investigate several tunings of these parameters in more detail in the following section.

\section{Phase diagrams} \label{sec: phase diagrams}

The overview of the previous section gives us a broad picture of how the phases vary for different values of the parameters. For further discussion, we now examine more closely the behavior for several particular parameter values which result in distinct characteristic patterns of the phase diagram vis a vis the relation of polyamorphism to glass physics. In particular, each material parameter tuning chosen below provides a different ratio of $T_c^{LL}/T_g$, describing the proximity of glass physics to polyamorphism. We discuss two materials further in the supplemental information. In the first additional material, the putative liquid-liquid phase transition would occur entirely below the Kauzmann transition, and thus there would not be, in fact, an equilibrium liquid-liquid phase transition within the present approximations. The second material arises from considering a much larger bond strength $\epsilon_b/\epsilon_{uu}=50$, similarly to what would be expected in liquid silica, which interacts via covalent bonds which are stronger than the hydrogen bonds of liquid water.

\begin{table}[h!]
    \centering
    \begin{tabular}[c]{|c|c|c|c|c|c|}
    \hline
        Section & $d_b$ & $\epsilon_{b}$ & $\epsilon_{uu}$ & $\epsilon_{bu}$ & $\epsilon_{bb}$ \\
        \hline
        \hline
        \ref{ssec: eq_crit}     & 1.2 & 7 & 1 & 0 & 1\\
        \hline
        \ref{ssec: water_crit}  & 1.2 & 7 & 1 & 0.75 & 1\\
        \hline
        \ref{ssec: Tg_crit}     & 1.2 & 7 & 1 & 0.9 & 1\\
        \hline
        SI                      & 1.2 & 7 & 1 & 1   & 1\\
        \hline
        SI                      & 1.2 & 50 & 1 & -1   & 1\\
        \hline

    \end{tabular}
    \centering
    \caption{Parameters of materials studied in section \ref{sec: phase diagrams} and in the supplemental information}\label{tab: params}
\end{table}

\subsection{Glass transition in the strongly 1st order regime, with clearly separable critical fluctuations}\label{ssec: eq_crit}
\begin{figure*}
    \centering
    \includegraphics[width=\linewidth]{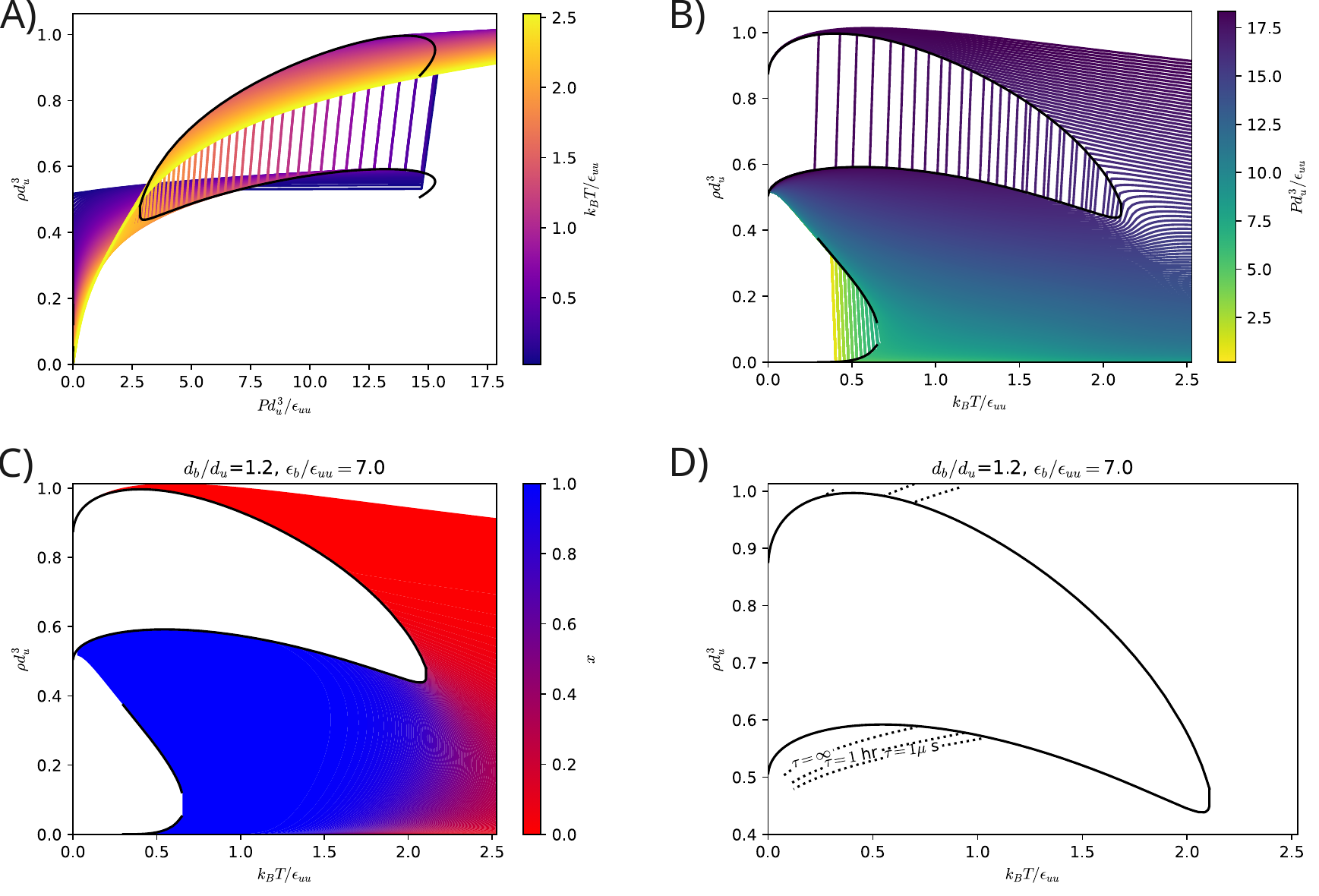}

    \caption{Here, we summarize the behavior of the model when $\epsilon_{ub} = 0$. This results in a liquid-liquid critical point which is well above the glass transition and thus accessible at equilibrium under reasonable timescales. (a) $\rho$ vs $P$ isotherms, where the color indicates the temperature. There is clear evidence of a liquid-liquid transition. A liquid-gas transition is also present, but only for very low densities. (b) $\rho$ vs T isobars, where the color indicates the pressure. We can see density maxima and minima near the critical point, where the density first decreases, then increases, then decreases again upon heating. Note here that the liquid-gas transition is clearly visible in the lower left. (c) Fraction of particles which are non-bonding, $x$, plotted as a function of $\rho$ and $T$, where the color indicates the $x$ value. Except at and above temperatures comparable to the critical temperature, $x$ is nearly exactly 0 (solid red) or 1 (solid blue). (d) Lines of constant configurational entropy, corresponding to the string crossover at $s_c=1.28 k_B$, the laboratory glass transition at $s_c = 0.79 k_B$, and the Kauzmann transition at $s_c = 0$.}
    \label{fig: eq_crit}
\end{figure*}

We first focus on a material where $T_c^{LL}$ is substantially higher than $T_g$. In this case, near $T_c^{LL}$ the critical effects can be completely separated from glassy dynamics, leading to a strong separation between the two scaling behaviors. The behavior of the liquid(s) near the critical point is well determined by the theory of ordinary phase transitions (where here of course the critical exponents take on their mean field values). The surface tension between the two phases vanishes, and other order parameters such as the density difference also vanish. Corrections to mean field are well understood through field theoretic and renormalization group approaches but here we neglect such effects. In contrast, near the glass transition temperature $T_g$, the critical fluctuations become nearly irrelevant, and phase fluctuations occur on smaller length scales in this regime. The surface tension varies slowly with temperature, and the dynamics of transitions between the polyamorphs are well described by classical nucleation theory with a correction for the timescale of glassy rearrangements \cite{greet_glass_1967,stevenson_ultimate_2011}. 

Figure \ref{fig: eq_crit} (a) shows the equilibrium density as a function of pressure, holding temperature fixed, figure \ref{fig: eq_crit} (b) shows the equilibrium density as a function of temperature, holding pressure fixed, figure \ref{fig: eq_crit} (c) the fraction of particles forming bonds $x$ as a function of density and temperature, and figure \ref{fig: eq_crit} (d) curves of constant configurational entropy and rearrangement timescales. We see a clear liquid liquid-phase separation with a critical point at $T_c^{LL}/\epsilon_{uu} = 2.1$. The two phases are both relatively dense, with a low density phase having a range $0.4\lesssim\rho d_u^3\lesssim0.6$ and a high density phase with a range $0.4\lesssim\rho d_u^3\lesssim1$. As one tunes either pressure (a) or temperature (b) across the phase transition, one sees discontinuous changes in the density, a signature of a first-order phase transition. Near $T_c^{LL}$, the densities of the two phases merge to a single critical density. We similarly see in figure \ref{fig: eq_crit} (b) and (c) that there is also a separate liquid-gas phase transition. The liquid-gas phase transition occurs only at very low pressures, and thus is difficult to see in figure \ref{fig: eq_crit} (a).  Figure \ref{fig: eq_crit} (c) shows the numerical value of $x$ remains roughly constant within each liquid phase, except very near the critical point. The high density liquid consists nearly of only non-bonded particles, while the low density liquid has nearly only fully bonded particles. As the critical point is approached within the low density phase, some of the bonds break, and in the supercritical regime for $\rho d_u^3\lesssim 0.5$, we see intermediate bond numbers. Figure \ref{fig: eq_crit} (d) is reproduced from figure \ref{fig: Tc, Tg, Tk} (b).

\subsection{Water-like critical point moderately higher than the glass transition} \label{ssec: water_crit}

\begin{figure*}
    \centering
    \includegraphics[width=\linewidth]{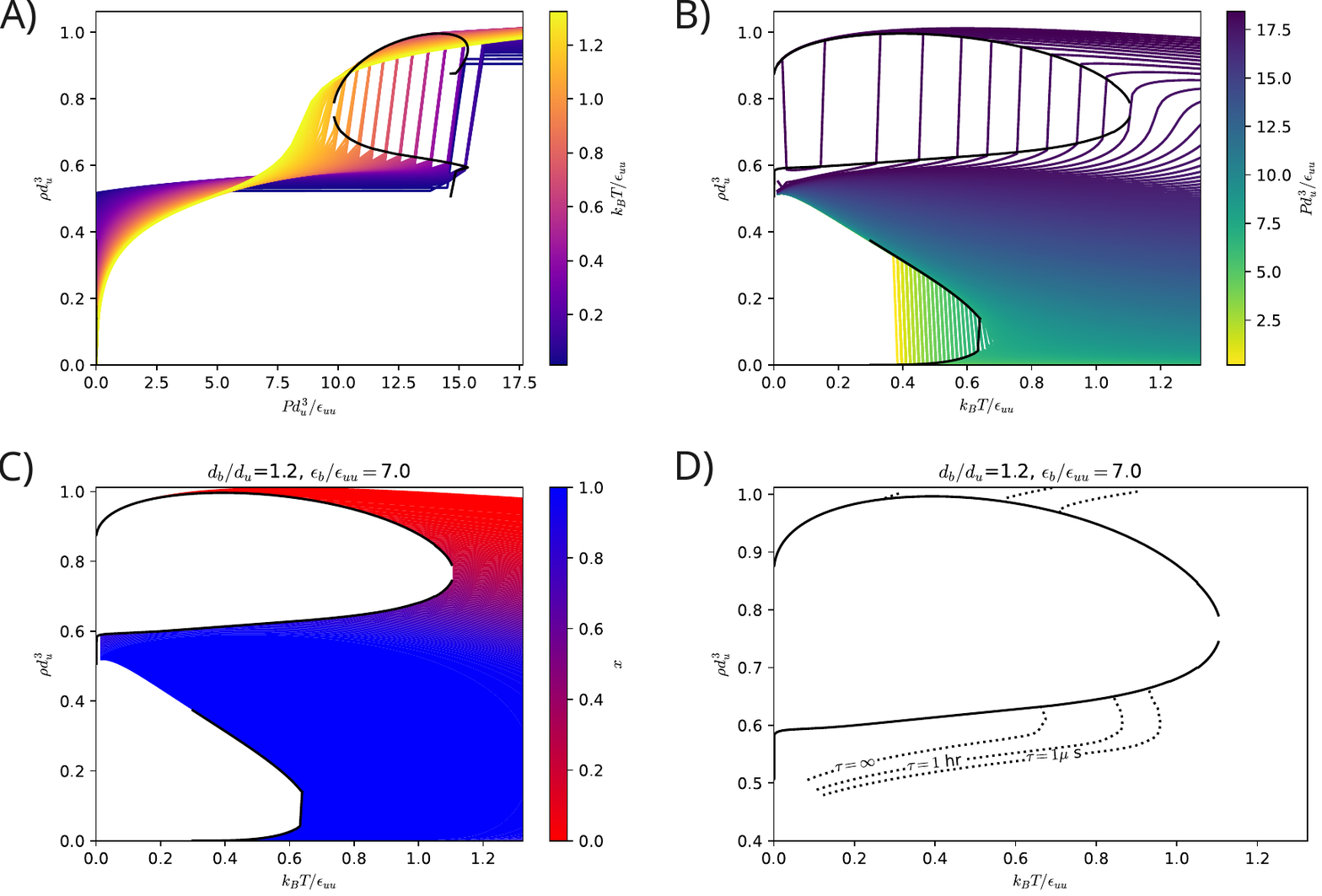}

    \caption{Here, we summarize the behavior of the model when $\epsilon_{ub}/\epsilon_{uu} = 0.75$. This results in a liquid-liquid critical point which is slightly above the glass transition. This critical point also lies in the ``no-mans' land'' where crystallization occurs before measurements can be taken. (a) $\rho$ vs $P$ isotherms, where the color indicates the temperature. Ignoring crystallization, there would be clear evidence of a liquid-liquid transition. A liquid-gas transition is also present, but only for very low densities. (b) $\rho$ vs T isobars, where the color indicates the pressure. We can see density maxima and minima near the critical point, where the density first decreases, then increases, then decreases again upon heating. Note here that the liquid-gas transition is clearly visible in the lower left. (c) The fraction of particles which are non-bonding, $x$, plotted as a function of $\rho$ and $T$, where the color indicates the $x$ value. Except at and above temperatures comparable to the critical temperature, $x$ is nearly exactly 0 (solid red) or 1 (solid blue). (d) Lines of constant configurational entropy, corresponding to the string crossover at $s_c=1.28 k_B$, the laboratory glass transition at $s_c = 0.79 k_B$, and the Kauzmann transition at $s_c = 0$. Notably, in the LDL, the lines of constant configurational entropy turn sharply at higher pressures and densities, where $x$ starts to decrease from 1. This indicates that bond breaking near the critical point can play a strong role in depressing the glass transition temperature.}\label{fig: water_crit}
\end{figure*}

Increasing the value of $\epsilon_{ub}$ corresponds to reducing the effective frustration between bonding and nonbonding particles. As $\epsilon_{ub}$ incraeses, the critical temperature for the liquid-liquid phase transition decreases. We note that the liquid-gas critical temperature and pressure remain largely unchanged with increasing $\epsilon_{ub}$, as do also the glass transition temperatures in both liquid phases. At a value of $\epsilon_{ub}/\epsilon_{uu} = 0.75$, the liquid-liquid critical point becomes roughly 30-40\% higher than the glass transition temperature in either liquid phase near the coexistence curve. This $T_c^{LL}/T_g$ ratio is comparable to what has been inferred for water \cite{velikov_glass_2001, poole_density_2005}. Notably, in many materials, this temperature range moderately above $T_g$ is also the no-man's land against crystal nucleation, where nucleation would occur at its highest rate according to a Turnbull analysis. In real water, it is thus very difficult to directly probe the liquid-liquid critical point in experiments, because crystallization occurs too quickly. We will discuss quantitatively the question of crystal stability and nucleation in the supplemental information.

We remind the reader however, that the density ratio $\rho_{HDL}/\rho_{LDL}$ found in the present model is somewhat larger than what is seen in water. In principle, a still more careful tuning of $d_b/d_u$ and $\epsilon_{bb}/\epsilon_{uu}$ could provide a still closer mimic for water. Indeed several computational models \cite{molinero_water_2009, stillinger_improved_1974} have been developed which more closely describe the exact behavior of water. Our purpose in the present paper is not, however, to precisely model particular laboratory materials, but rather to examine a model where both a liquid-liquid phase transition and a random first order transition play off each other.

Many aspects of the phase diagrams for the water-like case in figure \ref{fig: water_crit} are similar to those for the material described by the values of the parameter we have just discussed in the last section, shown in figure \ref{fig: eq_crit}. Notably, there are still two clear phase transitions, a liquid-liquid phase transition and a liquid-gas phase transition at lower pressures. The fraction of particles which form bonds, $x$ shown in figure \ref{fig: water_crit} (c) is again, roughly constant within each of the two liquid phases. Nevertheless, over a broader range of the low density liquid as this fraction changes on approaching the coexistence curve, where we see some of the bonds break. Since $T_g$ is relatively close to $T_c^{LL}$, we see a range of densities where only some of the possible bonds are formed near $T_g$. This results in a turnover in the curves of constant rearrangement timescale so that $T_g$ becomes a nonmonotonic function of pressure. The nonmonotonic $T_g$ value results from thermodynamic anomalies analogous to water's density maxima.

\subsection{Critical point near the glass transition} \label{ssec: Tg_crit}

\begin{figure*}
    \centering
    \includegraphics[width=\linewidth]{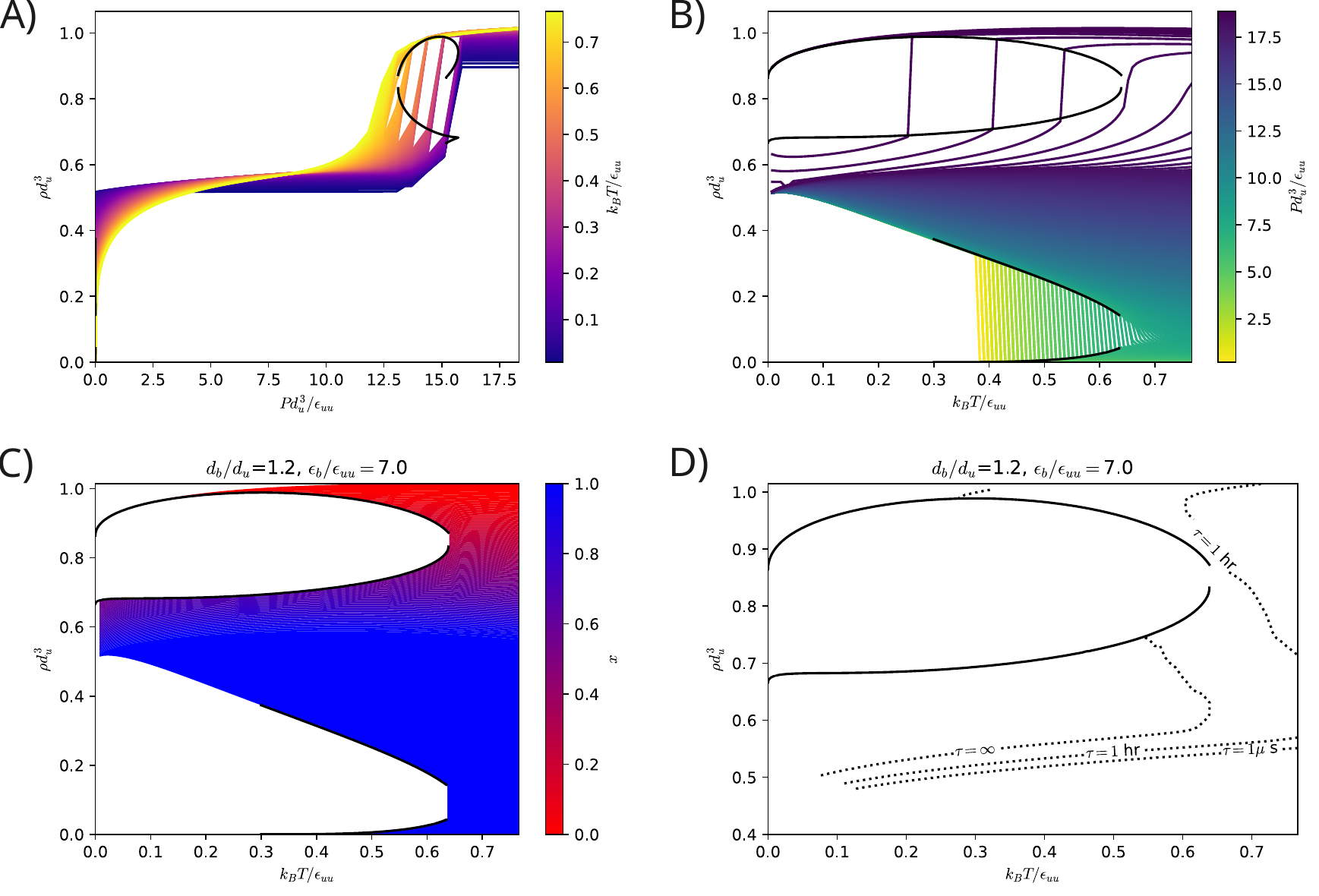}

    \caption{Here, we summarize the behavior of the model when $\epsilon_{ub}/\epsilon_{uu} = 0.9$. This parameter choice would result in a liquid-liquid critical point very close to the laboratory glass transition. In such a material, it would be difficult experimentally to directly demonstrate the liquid-liquid phase transition, but signatures of the polyamorphic phase transition still remain. (a) $\rho$ vs $P$ isotherms, where the color indicates the temperature. There is clear evidence of a liquid-liquid transition. A liquid-gas transition is also present, but only for very low densities. (b) $\rho$ vs T isobars, where the color indicates the pressure. We can see density maxima and minima near the critical point, where the density first decreases, then increases, then decreases again upon heating. Note here that the liquid-gas transition is clearly visible in the lower left. (c) The fraction of particles which are non-bonding, $x$, plotted as a function of $\rho$ and $T$, where the color indicates the $x$ value. Except at and above temperatures comparable to the critical temperature, $x$ is nearly exactly 0 (solid red) or 1 (solid blue). (d) Lines of constant configurational entropy, each corresponding to the string crossover at $s_c=1.28 k_B$, the laboratory glass transition at $s_c = 0.79 k_B$, and the Kauzmann transition at $s_c = 0$. The $T_g$ curve extends continuously from an LDL-like regime to a HDL-like regime as pressure is increased. $T_g$ is substantially non-monotonic as a function of $P$, indicating the presence of an underlying, inaccessibly liquid-liquid phase transition.}\label{fig: Tg_crit}
\end{figure*}

As we now continue decreasing the frustration between bonded and nonbonded particles, $T_c^{LL}$ decreases further. For $\epsilon_{ub}/\epsilon_{uu} = 0.9$, the equilibrium liquid-liquid phase transition would only occur now deep in the free energy landscape. Near $T_c^{LL}$, the calculated rearrangement timescales are around 1 hr. In a real material, such a liquid-liquid phase diagram would be difficult to probe directly. The ordinary approach for detecting a first-order phase transition is to look for a discontinuous change in density upon changing pressure or temperature. In this case, as shown in figure \ref{fig: Tg_crit}, the entire liquid-liquid phase transition occurs at temperatures and densities where the rearrangement timescale is longer than an hour. As such, any experiment would need to be performed extremely slowly (at least longer than the hour timescale) to allow the liquid itself to equilibrate. Alternatively, as is done in real glassy water measurements, the pressurization protocol may be performed on an out-of-equilibrium glass. However, discontinuous changes in the glassy phase are smeared out by fluctuations in configurational entropy, as the different mosaic states will undergo a density change at different pressures. We plan to investigate the role of glassy aging on the kinetics of this sort of nucleation further in a future work.

We stress additionally, that a full theory of the glass transition of the material presented in figure \ref{fig: Tg_crit} requires the inclusion of critical fluctuations. There are at least three length scales that are relevant for determining the dynamics of a material where the liquid-liquid critical point lies near $T_g$: the length scale for glassy rearrangements, the radius of a critical nucleation droplet, and the correlation length for critical fluctuations. A full theory which incorporates all three length scales is outside the scope of the present paper.

\section{Nucleation of polyamorphs} \label{sec: nucleation}

Now that we have established the equilibrium thermodynamic properties of the model for various parameter choices, we turn to the kinetics of several microscopic processes which may occur in polyamorphous network materials. We follow a scheme similar to that of Stevenson and Wolynes \cite{stevenson_ultimate_2011}, who analyzed the mechanisms of (crystal) nucleation in a deeply supercooled liquid. Several key takeaways will be important for our discussion, and we review them here.

First, each relevant kinetic process is governed by a distinct nucleation free energy profile $\Delta F_x(N) = N\Delta g_x+\sigma_x N^{y_x}$. The size of the nucleus at the free energy barrier, $N^\ddag$, and the barrier height $\Delta F^\ddag = \Delta F(N^\ddag)$ can be deduced easily by the condition $\partial \Delta F/\partial N = 0$. Similarly, the nucleus size, $N^*$, at which growth takes over can be deduced by the condition $\Delta F(N^*) = 0$. For glassy rearrangements within a liquid phase, $N^*$ is also the ultimate size of a rearrangement event, but of course neighboring regions can also reconfigure leading to dynamic facilitation effects \cite{xia_microscopic_2001, bhattacharyya_facilitation_2008}.

We have seen already that rearrangements in a supercooled liquid can be viewed as a nucleation of a liquid-like droplet surrounded by a glassy state, leading to $\Delta g_\alpha = -Ts_c$, $\sigma_\alpha = k_B T\gamma_{XW}$, and $y_\alpha = 1/2$. Rearrangements occur on a timescale $\tau_\alpha = \tau_0 \exp(\Delta g_\alpha^\ddag/k_B T)$, leading to the famous Vogel-Fulcher-Tamman law. The prefactor of this rate is a microscopic vibrational timescale on the order of $10^{-12}$ s.

Nucleation from a liquid to another thermodynamically distinct phase can occur through different routes, depending on an interplay of length scales. If the size of a glassy rearranging droplet, $N_\alpha^{*}$ is smaller than the nucleation region $N_{nucl}^\ddag$ (as is the case close to the phase transition, where $N_{nucl}^\ddag$ is large), then the nucleation event can be viewed as a transition from the equilibrium liquid, and $\Delta g_{nucl} = g_f - g_{l, i}$ where $g_f$ is the free energy of the final state (either a crystal as considered by Stevenson and Wolynes \cite{stevenson_ultimate_2011}, or a liquid as considered here) and $g_{l, i}$ is the entire liquid state free energy of the initial liquid state, including the configurational entropy of the initial liquid state. Nucleation is opposed by a surface tension with $y_{nucl} = 2/3$, as is typical on a non-wetted surface. In this regime, when $N_\alpha^{*}<N_{nucl}^\ddag$, so that cooperatively rearranging regions are smaller than the macroscopic nucleation droplet, classical nucleation theory is valid, and nucleation occurs on a timescale $\tau_{nucl} = \tau_{\alpha} \exp(\Delta g_{nucl}^\ddag/k_B T)$. The prefactor to the transition state activation factor is $\tau_\alpha$, reflecting a Kramers theory correction for dynamical recrossing events. Notably, upon cooling, the interplay of the two exponentially large factors leads to there being a temperature where the nucleation rate reaches a maximum, and upon further cooling further reductions in the nucleation barrier are more than balanced by increases in the rearrangement barrier reflected in the prefactor. The temperature regime of maximal nucleation rate for crystallization can sometimes lead to a so called ``no-mans' land'' if experiments on the liquid are inevitably disrupted by nucleation on laboratory time scales. The rate of nucleation occurring anywhere in a macroscopic sample during an experiment is also proportional to the physical size of the sample, because a larger sample has more volume in which individual nucleation events may occur. Owing to this, to avoid crystallization in poor glass formers, such as water \cite{mayer_new_1985} or metallic glasses \cite{meyer_cooling_2012}, experiments are sometimes performed on extremely small, aerosolized droplets of a liquid. In such experiments, crystals are unlikely to form due to the small system size.

If instead, $N_\alpha^{*}>N_{nucl}^\ddag$, then the liquid state does not sample distinct glassy structures during a nucleation event but initially is trapped in only one specific local structure. In this case we would see nucleation directly from the initially glassy state to the final (crystal or liquid) state, driven by a free energy difference $\Delta g_{nano} = g_f - g_{g, i} = \Delta g_{nucl}-T s_c$. This type of ``nanonucleation" will be spatially varying due to the fluctuations in $s_c$. In this regime, the Kramers' theory correction of the usual nucleation theory no longer applies. For supercooled liquids nucleating to crystals, nanocrystallization becomes important only quite a bit below the typical laboratory $T_g$. Stevenson and Wolynes \cite{stevenson_ultimate_2011} explained that heterogeneous nanocrystallization at a pre-existing crystal surface explains the otherwise puzzling transition to fast crystal growth that begins near $T_g$ \cite{kirkpatrick_crystal_1975} for some substances. Signatures of bulk homogeneous phase nanocrystallization are not seen in most typical laboratory glass experiments, but they can be probed instead by geological and ultrastable glasses. Stevenson and Wolynes suggest that there are two possible outcomes: one at extremely low temperatures, where $N^*_{\alpha}>N^*_{nano}>N^\ddag_{nucl}$, and consequently the nanocrystallite (or nanodroplet) has no tendency to disappear due to glassy rearrangements. Once the nanocrystallite (or nanodroplet) forms, it grows in a way analogous to bulk crystallization, although with different nucleation kinetics due to the different nucleation mechanism. This process is known as ``direct nanonucleation". At intermediate temperatures, where in contrast $N^*_{\alpha}>N^\ddag_{nucl}>N^*_{nano}$, the resulting structures depend on an interplay of the length- and timescales of nucleation, growth, and glassy rearrangement. Depending on the fluctuations in $s_c$, certain ``mosaic cells'' of the liquid would nanonucleate more easily than others. If the fraction $p$ of such cells exceeds the percolation threshold, then a large scale percolating network of nanocrystals (or nanodroplets) will form through a relatively fast heterogenous nucleation mechanism. This process is known as ``percolative nanonucleation". For crystallization, Orrit has shown the existence of this percolative structure in his studies of the mechanical properties of glycerol \cite{yuan_communication_2012}. In either form of nanonucleation, the simple kinetic picture of Turnbull's classical nucleation theory breaks down, and the heterogeneous glassy nature of the liquid must be explicitly considered in order to construct a complete picture of the dynamics of the nucleation process.

In the sections that follow, the time and length scales of the several different kinetic processes, including cooperative rearrangement, macroscopic nucleation, and nanonucleation, will be calculated individually and compared. A natural quantity to compare is the overall free energy barrier for a process to occur, including the contribution of any exponentially large prefactors, such as the dynamic prefactor of the bulk nucleation rate. We note, however, that while relaxation times due to cooperative rearrangement are typically reported as a timescale, nucleation rates are correctly understood in terms of the rate of crystal nucleation events per unit time per unit volume. A larger sample of a material has a greater chance for nucleation to occur simply because there are more locations where the nucleation event may take place before the nucleus of the new phase later grows to fill the container. Implicit to our comparisons, which rely first and foremost on the free energy barriers, is that we are considering the timescale for rearrangement or nucleation to occur within some particular microscopic region of volume determined by the relevant length scale. For both cooperative rearrangement and nucleation, we report a timescale on which the kinetic process will occur around a particular given particle in the sample. In contrast, the timescale for nanonucleation plotted in the figure is the timescale on which nanonucleation will occur in a particular given mosaic cell - a region the size of the rearranging droplet. This means that the relative rate for nanonucleation events to occur compared to bulk nucleation occurring anywhere in a large sample is $N_{\alpha}^\ast \tau_{bulk}/\tau_{nano}$, a factor of $N_{\alpha}^\ast$ larger than would be na\"ively inferred based on the free energy barriers. $N_{\alpha}^\ast$ is at most a few hundred particles near the glass transition, while the kinetic timescales vary exponentially, so this size correction is modest.

\subsection{Surface tensions}

\begin{figure*}
    \centering
    \includegraphics[width=\linewidth]{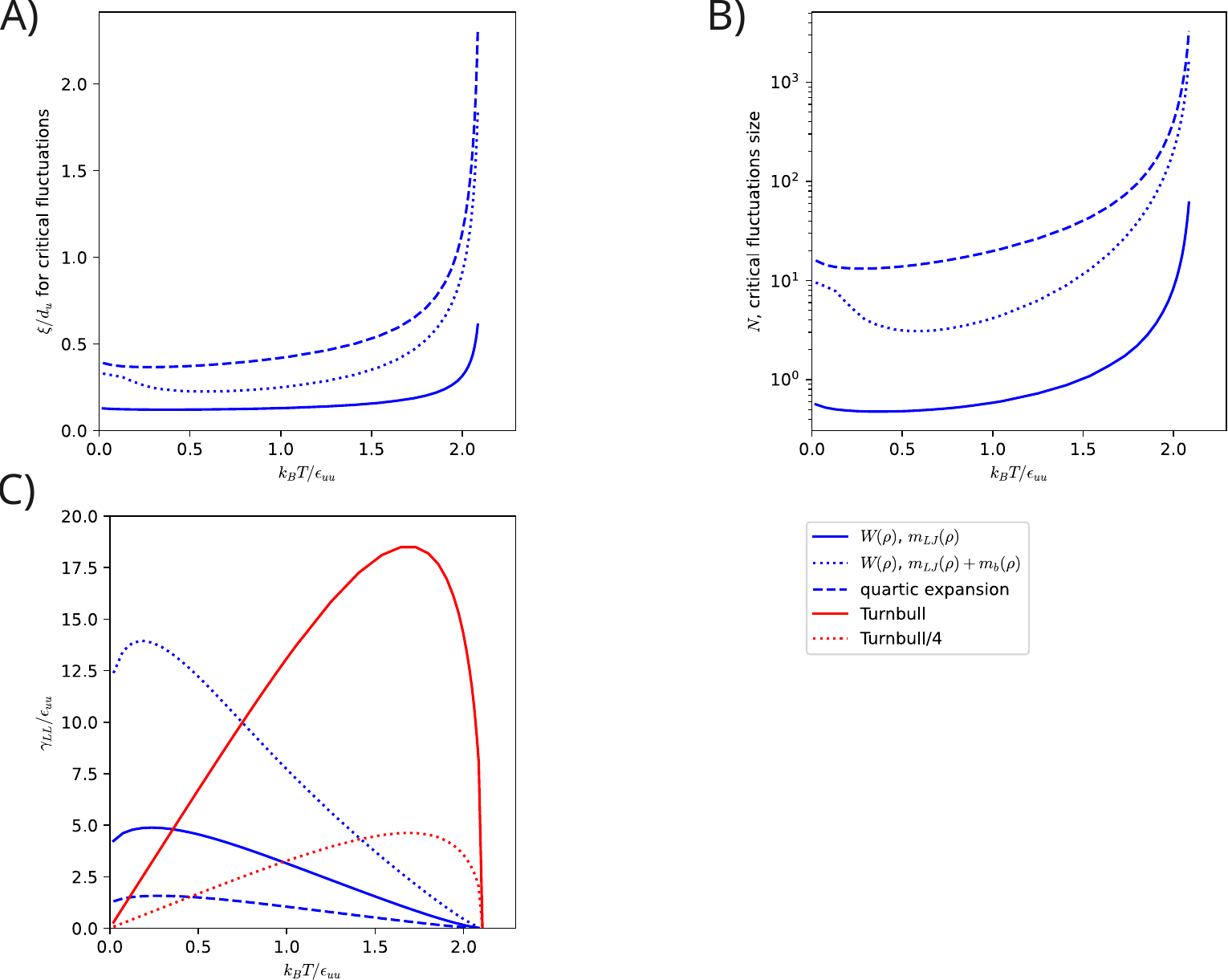}

    \caption{Here, we describe the protocol for calculating the surface tensions between liquid phases. (a) First, a Landau-Ginzberg like expansion around the critical point, along with data from the coexistence curve, is used to compute the length scale of critical fluctuations. This length scale diverges near the critical point, and quickly decays to a small, finite value. The critical correlation length is related to the interface thickness of a surface between the two liquid phases, which is around 5 particle diameters thick in this case. (b) Similar calculations are used to find the surface tension near the critical point, where the length scale of fluctuations is much larger than the particle spacing. Far from the critical point, it is instead more accurate to use a thin interface approximation due to Turnbull \cite{turnbull_formation_1950} which estimates the surface tension based on the enthalpy difference between the two phases .}\label{fig: surface tensions}
\end{figure*}
Just as for crystallization \cite{stevenson_ultimate_2011}, in order to quantify nucleation processes from one liquid state to the other liquid form, we need to know the surface tension between the bulk phases. Various estimations can be made. In general, the surface tension between two macroscopic phases should vanish at the critical point, where the two phases become identical. This critical situation can not take place for three dimensional crystallization due to the symmetry breaking inherent to crystal melting. Generally the surface tension will be higher for two phases that are more structurally or thermodynamically different from one another. In this paper we will consider several methods of estimating the surface tension between the bulk liquid phases, and stress that each estimate is valid under different conditions. The various approximate surface tension estimates we have considered are described below.

One estimate, $\gamma_{int}$, a field theoretic result from liquid state theory, is described very well in the book by Widom and Rowlinson \cite{rowlinson_molecular_2013}. The surface tension is related to $W(\rho)$, the negative excess Gibbs free energy per volume; $m(\rho)$, the strength of the local attractive forces; and implicitly $\xi(\rho)$, the density profile at the interface. We will consider two formulas for $m(\rho)$, one which includes only the nonbonding interactions and one which additionally includes the bond strength as an attractive force.

The field theoretic results are most accurate near the critical point, where the interface thickness is at least as wide as several particle diameters. This continuum limit allows one to determine the surface tension as an integral over a freely varying density profile, which involves balancing the excess Gibbs free energy density $-W(\rho)$ against a square gradient term depending on the strength of intermolecular attractions $m(\rho)$, which may be calculated from the intermolecular potential
\begin{gather}
    m(\rho) = -\int_0^\infty d\bm{R} R^2u(\bm{R})
\end{gather}
where $u(\bm{R})$ is the attractive part of the intermolecular potential. For the nonbonding forces, we use the same splitting of the potential as before, due to Weeks, Chandler and Andersen. For such nonbonding forces, one finds
\begin{gather}
    m_{LJ}(\rho) = \frac{13\pi}{21} 2^{11/6}\bigg(\epsilon_{uu}(1-x)^2 +2\epsilon_{ub}x(1-x)+\epsilon_{bb}x^2\bigg)
\end{gather}
There is an additional term coming from the short range bonding force
\begin{gather}
    m_b \approx 2x\epsilon \frac{4\pi d_b^5}{5}
\end{gather}
as long as $\sqrt{2\epsilon_b/\kappa_b}\ll d_b$. Once $m(\rho)$ is known, the density profile $\rho(z)$ as a function of the distance from the interface $z$ is determined from the Euler-Lagrange equations 
\begin{equation}
    m(\rho)\frac{d^2\rho}{dz^2} = -\frac{dW(\rho)}{dz}
\end{equation}

The surface tension along the coexistence curve follows directly from
\begin{equation}
    \sigma = \int_{\rho_1}^{\rho_2} \sqrt{-2m(\rho)W(\rho)} d\rho
\end{equation}
where $\rho_i$ are the coexistence densities and $W(\rho) = \rho \min_{x}\bigg[g_l(\rho_i, x_i) - g_l(\rho, x)\bigg]$. The length scale of the interface (and thus the length scale for critical fluctuations in the vicinity of the critical point) can be calculated from the density profile directly, whose inverse is 
\begin{equation}
    z(\rho) = \int\sqrt{-\frac{m(\rho)}{2W(\rho)}}d\rho
\end{equation}
For the sake of simplicity, we will assume the liquid-liquid surface tension does not change much when moving away from the coexistence curve. This is reasonable until the liquid approaches the mean field spinodal limit, at which point the interface itself becomes ramified and the surface tension drops to zero, much as near the critical point \cite{cahn_free_1959, unger_nucleation_1984}.

In addition to the surface tension estimate discussed above, we consider a surface tension $\gamma_{quart}$ which is based purely on a Landau-Ginzburg expansion of the microscopically determined Gibbs free energy to the fourth order around the critical point. his estimate approximates the free energy as a quartic polynomial, and relates the surface tension to derivatives of the free energy near the critical point. While this expansion has the benefit of computational simplicity, it loses accuracy away from the critical point. 

We also consider the surface tension estimate $\gamma_{Turnbull}$, which follows from an analog of Turnbull's rule for melting. Turnbull's rule works well for crystal melting, which is strongly first order. Here, the surface tension is proportional to $\Delta H$ at the transition point, or equivalently proportional to $T \Delta S$ at the transition point. Since the Turnbull rule is accurate for solid-liquid transitions (melting) and liquid-liquid surface tensions are known to be somewhat lower than crystal-liquid surface tensions, we also show results for an estimate where we have reduced the Turnbull result by a proportionality constant inferred from simulations \cite{wang_direct_2026}.

Within our model, a rich diversity of possible nucleation regimes can occur. For a particular route between the phases, we must specify a set of microscopic material parameters ($\epsilon_b$, $\epsilon_{XY}$, etc.), starting and ending phases, an approximation for the surface tension, and a pathway through $P$ and $T$ space. For simplicity, we discuss here nucleation of the polyamorphic transition with the parameters used in section \ref{sec: phase diagrams}, which have critical points ranging from well above $T_g$ to below $T_g$. We also restrict ourselves to simple pressure and temperature quenches, where either $T$ or $P$ is held fixed, and the other thermodynamic variable is changed in the quench.

\subsection{Pressurization through the phase transition}
\begin{figure*}
    \centering
    \includegraphics[width=\linewidth]{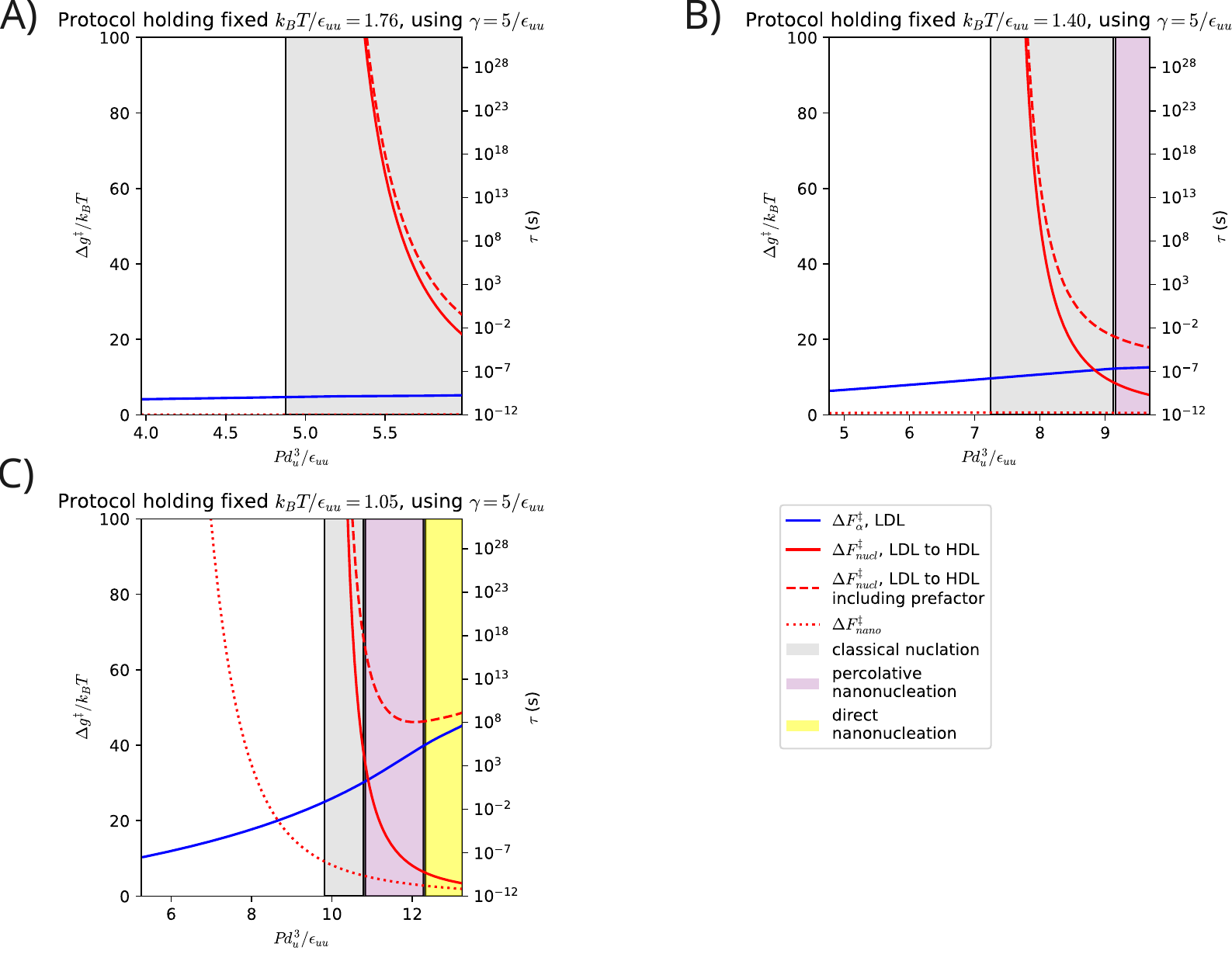}

    \caption{Here, we plot the free energy barriers and timescales for several kinetic processes which may occur along pressurization protocols across the phase diagram. Nucleation rates are reported based on the timescale for nucleation to occur in a particular microscopic region, where for bulk nucleation, this region is a given particle and for nanonucleation, this region is the size of the cooperatively rearranging region. The liquid-liquid surface tension is chosen to have a fixed value $\gamma/\epsilon_{uu} = 5 $. Pressurization is performed at three different temperatures, so that the liquid-liquid phase transition occurs (a) well above $T_C$, (b) near $T_C$, and (c) well below $T_C$. In each case, we plot the rearrangement barrier in solid blue and the (bulk) nucleation barrier in solid red. The total bulk nucleation rate depends on the sum of these barriers, shown in dashed red. The barrier to nanonucleation is shown in dotted red, although nanonucleation is only opposed by a finite barrier in (c). We see that in (a), where rearrangement rates remain relatively low, nucleation occurs entirely via classical nucleation. At the other extreme, in (c), the dynamic prefactor to nucleation prevents any bulk nucleation from occuring, and only nanonucleation can occur on laboratory timescales.}\label{fig: protocol pressurization, several T}
\end{figure*}
A common protocol used to investigate the liquid-liquid phase transition is as follows: a low density liquid is first directly cooled down to low temperatures. Of course in the laboratory (and sometimes in the computer), the liquid often crystallizes, but if it does not crystallize, the liquid will fall out of equilibrium and form a glass. After being cooled, the liquid is then pressurized until bonds break and a discontinuous drop in the volume would be observed. Pressurization protocols in amorphous glassy ice were among the first experiments to suggest the existence of a second liquid phase in water \cite{mishima_apparently_1985}. The results of this procedure depend on the quench temperature, as shown in figure \ref{fig: protocol pressurization, several T}. If the temperature is sufficiently high, glassy rearrangements are fast, and the liquid nucleates via a classical nucleation theory - like droplet. On the other hand, at temperatures around and below $T_g$, nucleation is limited by the timescale of rearrangement events. An ordinary nucleation event would not be observed on human timescales, and instead the phase transition would occur whenever the pressure reaches the limit of stability of the glass.

\subsection{Cooling through the phase transition}

\begin{figure*}
    \centering
    \includegraphics[width=\linewidth]{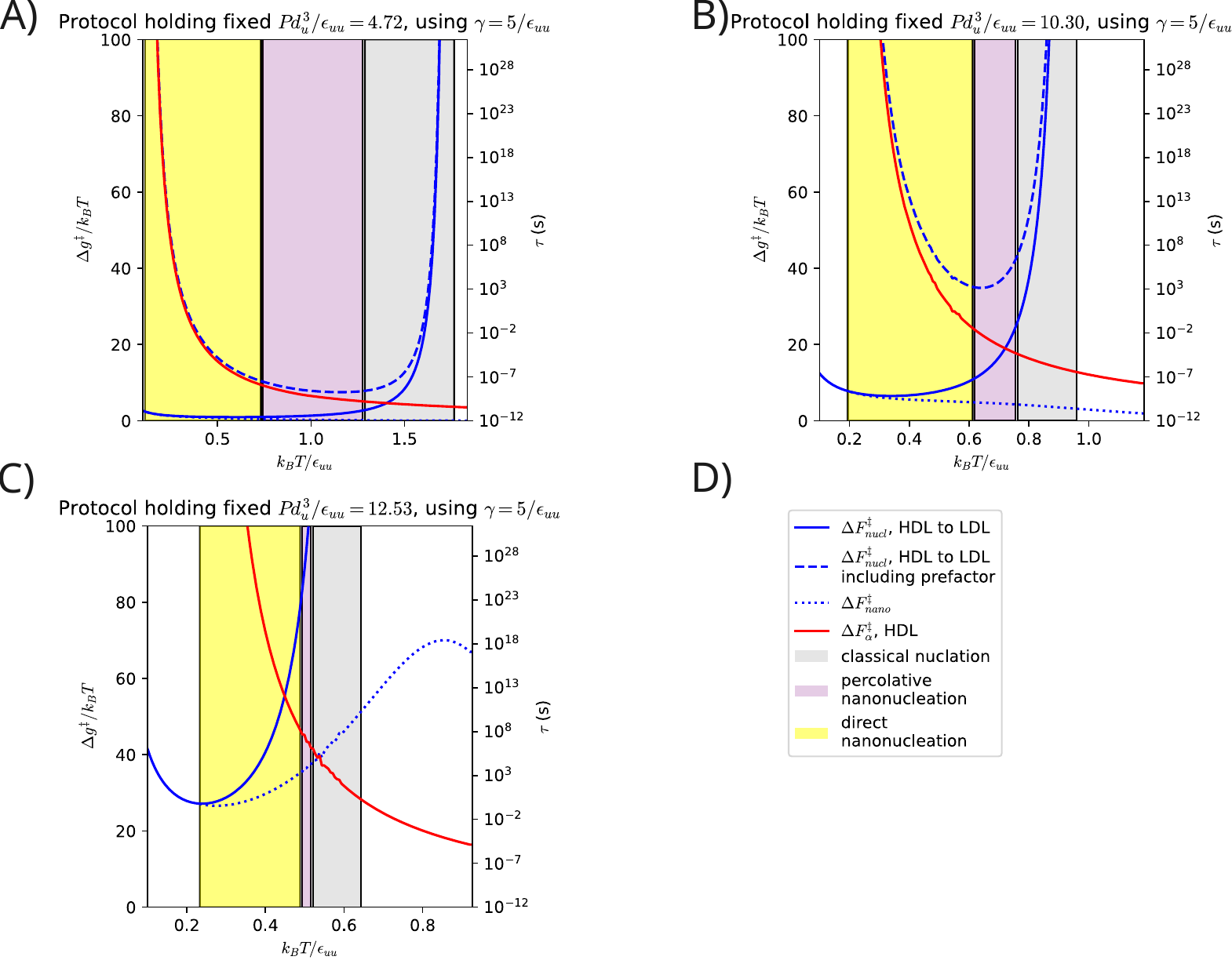}

    \caption{Here, we plot the free energy barriers and timescales for several kinetic processes which may occur along cooling protocols across the phase diagram. Nucleation rates are reported based on the timescale for nucleation to occur in a particular microscopic region, where for bulk nucleation, this region is a given particle and for nanonucleation, this region is the size of the cooperatively rearranging region. The liquid-liquid surface tension is chosen to have a fixed value $\gamma/\epsilon_{uu} = 5$. Cooling is performed at three different pressures, so that the liquid-liquid phase transition occurs (a) well above $T_C$, (b) near $T_C$, and (c) well below $T_C$. In each case, we plot the rearrangement barrier in solid red and the (bulk) nucleation barrier in solid blue. The total bulk nucleation rate depends on the sum of these barriers, shown in dashed blue. The barrier to nanonucleation is shown in dotted blue, although nanonucleation is not opposed by a finite barrier in (a), and is thus not clearly visible. We see that in (a), where rearrangement rates are relatively low near the phase transition, nucleation occurs very readily via classical nucleation and extremely rapid quenching is necessary to avoid bulk nucleation. At the other extreme, in (c), the dynamic prefactor to nucleation prevents any bulk nucleation from occurring, and in fact the total bulk nucleation timescale is above the scale timescale axis. In the intermediate regime (b) we have illustrated a case where nucleation proceeds at a minimum timescale of 1 hr, which is ideal for direct laboratory studies of glassy aspects of liquid-liquid nucleation.}\label{fig: protocol cooling, several P}
\end{figure*}

The phase transition between the two liquids may also be crossed by cooling, starting from the high density liquid, as shown in figure \ref{fig: protocol cooling, several P}. If cooling is performed at sufficiently low pressures, the liquid will cross the phase transition at a temperature $T_{LL}>T_g$. This again results in a classical nucleation theory - like regime, where nucleation occurs based on a nucleation free energy barrier with a dynamic prefactor based on the rearrangement timescale. If instead, the cooling is performed at higher pressures, the liquid would fall out of equilibrium (i.e. cross the glass transition) well before reaching the phase boundary, and thus nucleation as described by Turnbull does not occur on experimental timescales due to the $>1$ hr dynamic prefactor. We note however, that other routes for a phase change are conceivable, particularly near the spinodal for the liquid-liquid phase transition where the critical nuclei become ramified and the simple picture of a droplet-like nucleation event breaks down.

\subsection{Varying the surface tension}
As has approximately been done for melting \cite{stevenson_ultimate_2011}, the previous discussion in figure \ref{fig: protocol cooling, several P} assumes the polyamorphic surface tension varies only slowly with temperature. We must remember that an important distinction between liquid-liquid nucleation and liquid-solid nucleation, however, is that the liquid-liquid surface tension near a critical point varies much more strongly with temperature than does the crystal liquid interfacial tension. For transformations to both a solid and to a liquid, as temperatures is reduced, the driving force for nucleation gets larger and the driving force for rearrangement gets smaller. For the constant surface tension, this results in a decreasing nucleation barrier height competing with an increasing dynamic prefactor. While for the liquid to crystal transition, this is reasonable, the surface tension in a polyamorphic liquid-liquid transition varies more dramatically, especially near the critical point. Thus, we must worry about a third competing effect: namely, that upon cooling, the surface tension itself varies, in addition to the driving forces for nucleation and rearrangement.

\begin{figure*}
    \centering
    \includegraphics[width=\linewidth]{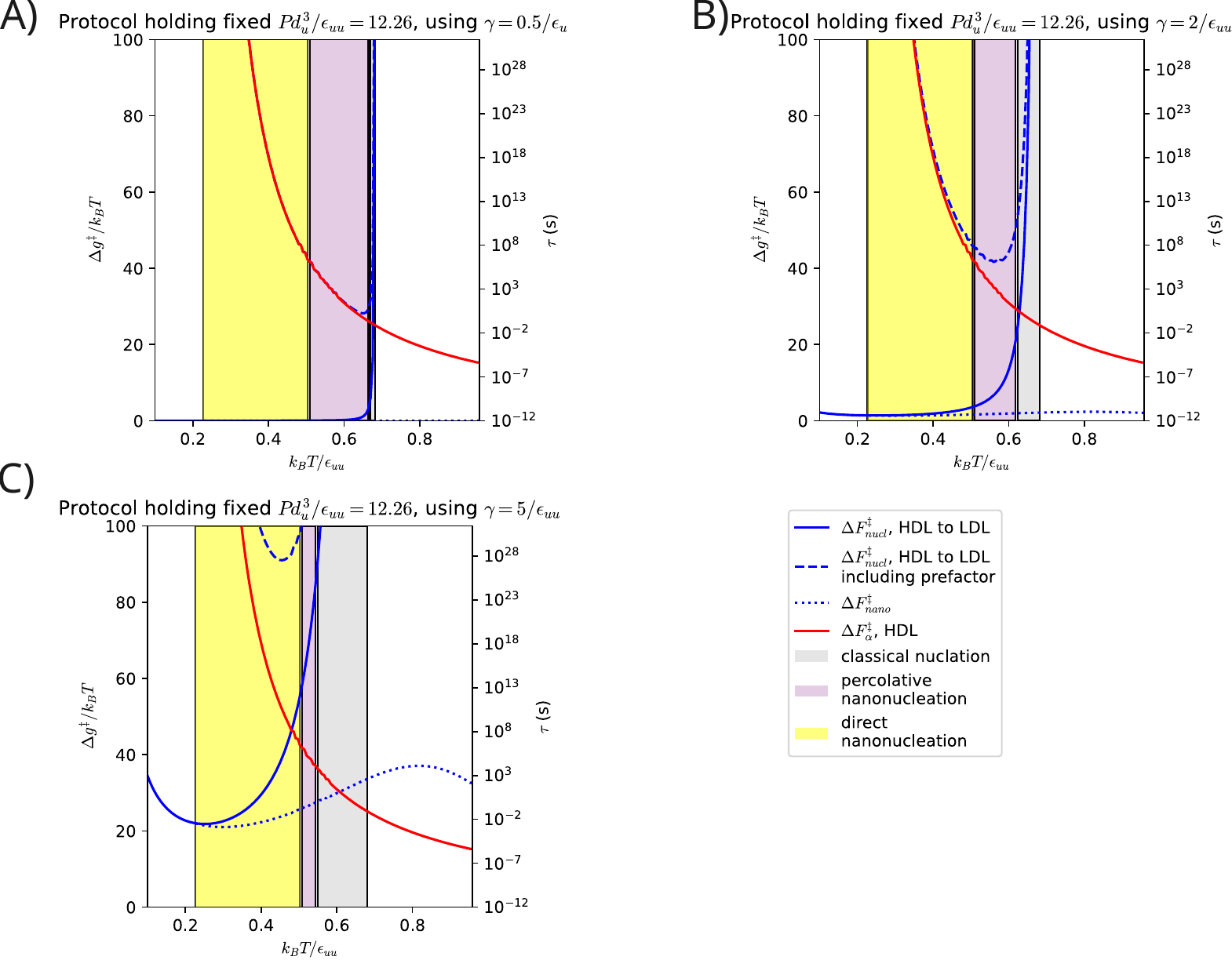}

    \caption{Here, we plot the free energy barriers and timescales for several kinetic processes which may occur along cooling protocols across the phase diagram. Nucleation rates are reported based on the timescale for nucleation to occur in a particular microscopic region, where for bulk nucleation, this region is a given particle and for nanonucleation, this region is the size of the cooperatively rearranging region. The liquid-liquid surface tension is chosen to have a fixed value (a) $\gamma/\epsilon_{uu} = 0.5$, (b) $\gamma/\epsilon_{uu} = 2$, and (c) $\gamma/\epsilon_{uu} = 5$. Cooling is performed at $P d_u^3/\epsilon_{uu} = 12.26$, which is chosen so that $T = T_C$ at the phase transition, resulting in a dynamic prefactor of 1 $\mu$s in the high density liquid at the coexistence point. In each case, we plot the rearrangement barrier in solid red and the (bulk) nucleation barrier in solid blue. The total bulk nucleation rate depends on the sum of these barriers, shown in dashed blue. The barrier to nanonucleation is shown in dotted blue, although nanonucleation is not opposed by a finite barrier in (a), and is thus not clearly visible. We see that in (a), for very low liquid-liquid surface tensions, the bulk nucleation timescale closely follows the rearrangement timescale, except for diverging in a small temperature range near the phase transition. The regime where nucleation occurs via a Turnbull-like process is extremely narrow. As the surface tension is increased, the classical nucleation regime grows and the percolative nanonucleation regime shrinks. At the same time, higher surface tensions naturally result in higher free energy barriers to both forms of nucleation due to the greater energetic cost to form a nucleation droplet.}\label{fig: protocol cooling, several gam}
\end{figure*}

To understand the role that a variable surface tension plays in nucleation of polyamorphic phases, we first examine what happens when we tune the surface tension by hand across a range of values, for the same cooling protocol in the same material, as shown in figure \ref{fig: protocol cooling, several gam}. For very low surface tensions, the nucleation timescale becomes completely dominated by the dynamic prefactor. After crossing the phase transition, the nucleation timescale drops rapidly to just above the rearrangement timescale. As the temperature is reduced, this timescale gradually increases until it diverges at $T_K$. On the other hand, for higher surface tensions, the bare nucleation barrier height is much larger.

\begin{figure*}
    \centering
    \includegraphics[width=\linewidth]{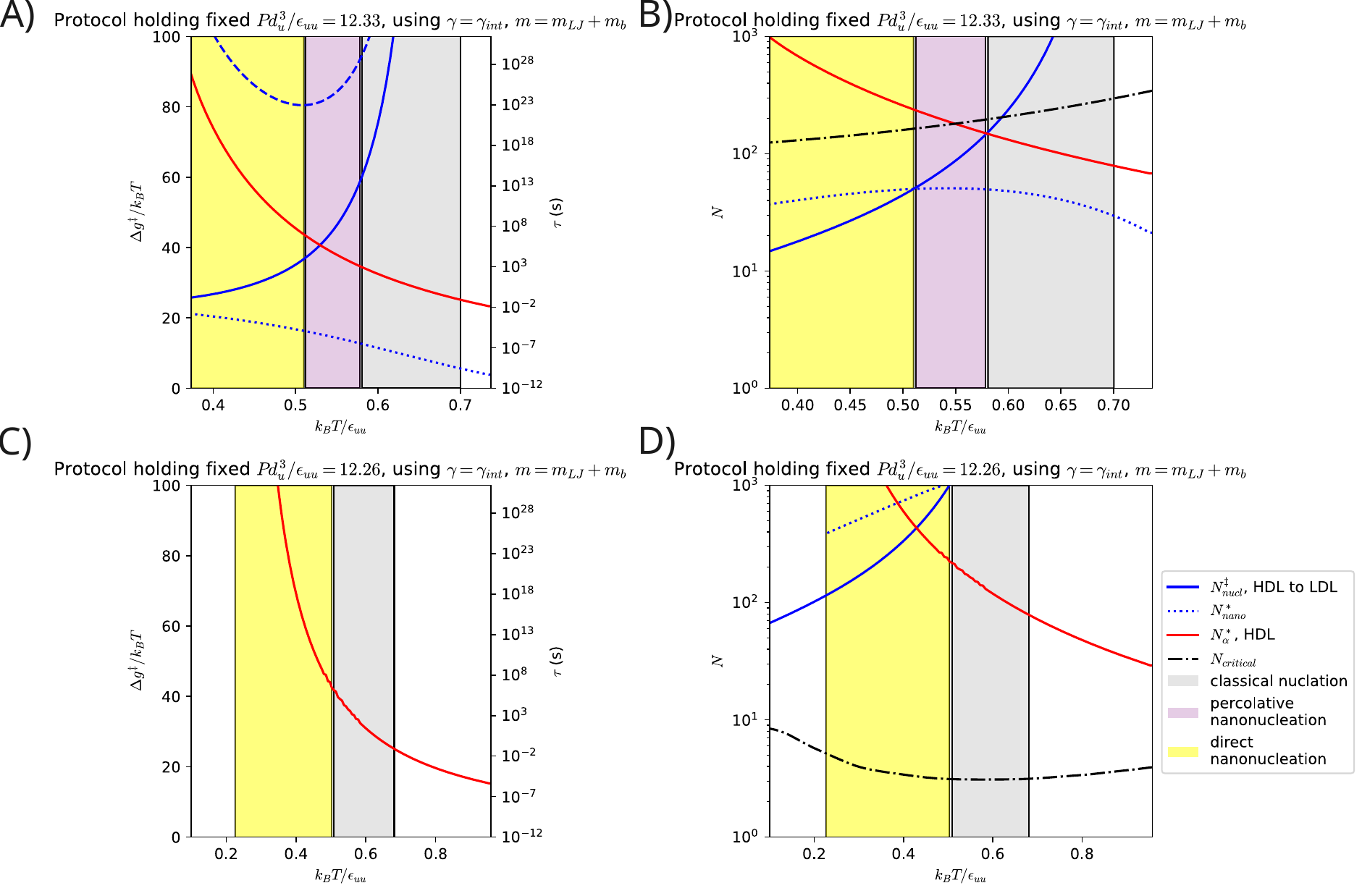}

    \caption{Here, we plot (a, c) the free energy barriers and timescales and (b, d) the number of particles involved for several kinetic processes which may occur along cooling protocols across the phase diagram. Nucleation rates are reported based on the timescale for nucleation to occur in a particular microscopic region, where for bulk nucleation, this region is a given particle and for nanonucleation, this region is the size of the cooperatively rearranging region. (a) and (b) correspond to the phase diagram with $\epsilon_{ub}/\epsilon_{uu} = 0.75$ and thus $T_c^{LL}/T_g\approx 1.4-1.5$ while (c) and (d) correspond to the phase diagram with $\epsilon_{ub} = 0$ and thus $T_c^{LL}/T_g\approx 1.8-2.0$. The liquid-liquid surface tension is calculated based on a field theoretic approach to the liquid which relates the surface tension to an integral over the excess Gibbs free energy density $-W(\rho)$ and $m(\rho)$, which captures the short-ranged interactions between particles in the droplet interface. Cooling is performed at (a-b) $P d_u^3/\epsilon_{uu} = 12.33$ or (c-d) $P d_u^3/\epsilon_{uu} = 12.26$, which are chosen so that $T = T_C$ at the phase transition, resulting in a dynamic prefactor of 1 $\mu$s in the high density liquid at the coexistence point. In each case, we plot the rearrangement barrier (or region size) in solid red and the (bulk) nucleation barrier (or droplet size) in solid blue. The total bulk nucleation rate depends on the sum of these barriers, shown in dashed blue. The barrier (or region size) to nanonucleation is shown in dotted blue, although nanonucleation is not opposed by a finite barrier in (c), and is thus not clearly visible. In (b) and (d), we also plot in black the number of particles involved in a critical fluctuation. We see that in (b), where $T_c^{LL}$ is relatively near to $T_g$, critical fluctuations are of substantial size, and in fact are larger than the rearrangement region throughout the classical nucleation regime. In contrast, in (c), where $T_c^{LL}$ is further from $T_g$, the critical fluctuations are much smaller and can be safely neglected from the theory.}\label{fig: protocol cooling, gam RW}
\end{figure*}

The variation of $\gamma(T)$ is most relevant to the rate of liquid-liquid nucleation when $T_c^{LL}$ is above $T_g$, but not by too far. The parameters for a water-like critical point give a good example of this. Three of our estimates for the surface tension provide the correct trend for the temperature dependence of the surface tension near $T_g$ and $T_C$, namely those surface tension estimates which derive from field theoretic integrals over a continuous density profile. In figure \ref{fig: protocol cooling, gam RW} we show the nucleation barriers for a protocol passing through $T_C$ using each of these surface tension estimates. 

A notable feature of the nucleation of the water-like material in figure \ref{fig: protocol cooling, gam RW}(a-b) is that the length scale of critical fluctuations is sizable compared to the length of both glassy rearrangements and nucleation clusters. This implies that the liquid-liquid interfaces would be substantial softened, and the true nucleation mechanism may occur via a route not studied here, where both the critical and the glassy properties of the liquids are important. Generally, both critical and glassy aspects are simultaneously important for cooling or pressurization protocols where $T\sim T_g\sim T_c^{LL}$. We show an exaggerated case of such a nucleation process in the supplemental information.

\section{Discussion}
In this paper, we have presented a simple microscopic model for liquid-liquid phase transitions to which the RFOT theory of glasses may be easily applied at the molecular level. By characterizing the glassy and liquid states of the model, we describe the interplay of the key thermodynamic and kinetic properties characteristic of network liquids relevant to glassy materials with underlying liquid-liquid phase transitions. Model liquids within this framework readily display water-like anomalies, including density maxima and minima and maxima in the heat capacity and compressibility. The tunable nature of the model allows one to study examples where the liquid-liquid critical point can be found either well above, nearby, or even below the glass transition, depending on the precise tuning of the molecular parameters. All of these possible regimes have been invoked by others to explain a variety of experimental systems. The nature of the nucleation dynamics from one liquid phase to the other depends in detail on this interplay of $T_g$ and $T_c^{LL}$. The liquid-liquid surface tension varies more strongly with supercooling for polyamorphic transitions than does the crystal-liquid surface tension due to the critical fluctuations that can occur for the liquid-liquid phase transition, which has a critical point. Owing to this, generally nanonucleation plays a greater role in the nucleation kinetics of a typical liquid-liquid phase transition than it does for the kinetics of crystallization, where nanonucleation is only directly manifested in bulk upon very strong supercooling. While we have located the regimes where the nanonucleation scenarios prevail, further study will be needed to quantify the role of critical fluctuations in setting the structural details of phase separated glasses formed in this way. The critical fluctuations must be accounted for in the kinetics of a liquid-liquid phase transition when the critical point is near the glass transition.

\begin{acknowledgments}
This research was supported by the Center for Theoretical Biological Physics sponsored by the NSF (Grants PHY-2019745). P.G.W. is also supported by the D.R. Bullard-Welch Chair at Rice University (Grant C-0016).
\end{acknowledgments}

\section*{Data Availability Statement}

Data sharing is not applicable to this article as no new data were created or analyzed in this study.
\section*{Conflict of Interest Statement}
The authors have no conflicts to disclose.

\bibliography{references}

\section{Supplemental Information}
\subsection{Supercooled critical point}

\begin{figure*}
    \centering
    \includegraphics[width=\linewidth]{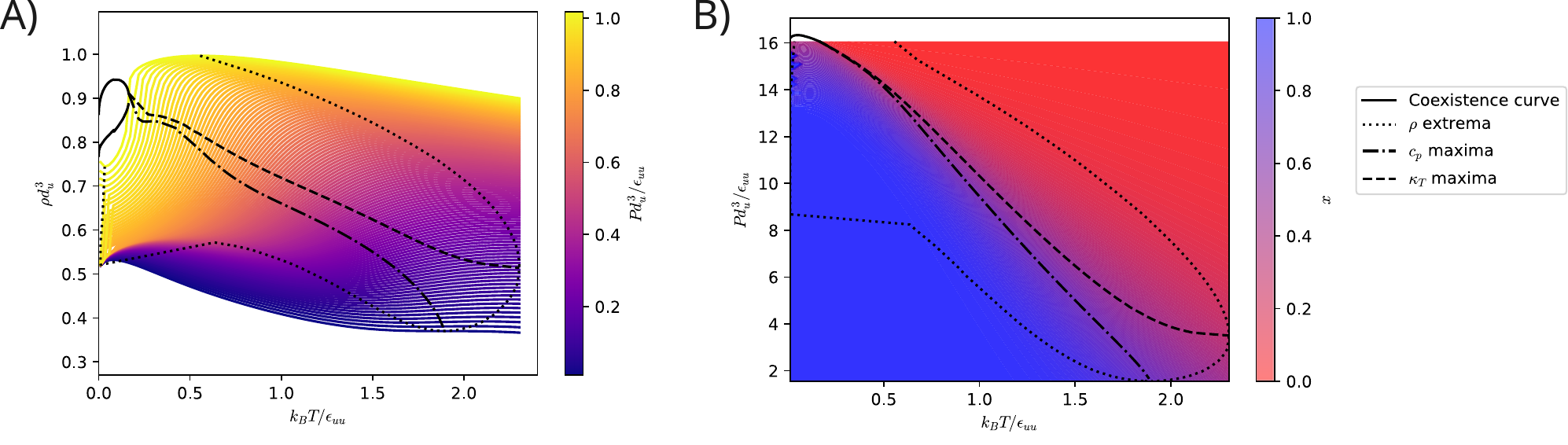}
    
    \caption{This plot is the analog of figure \ref{fig: density extrema}, for a material with $\epsilon_b/\epsilon_{uu} = 7, \epsilon_{ub}/\epsilon_{uu} = \epsilon_{bb}/\epsilon_{uu} = 1$. The $x$ dependence of the repulsive force in the equation of state provides a slight frustration between disparate local structures, which is sufficient for a liquid-liquid phase transition over a small range of temperatures. Despite the putative liquid-liquid phase transition occurring only below $T_K$ (and thus being aphysical in the context of an equilibrium liquid), it still mathematically results in density maxima maxima and other thermodynamics anomalies. We show density, compressibility, and heat capacity extrema with (a) corresponding isobars as a function of temperature and (b) bonding fraction, where we can see that the thermodynamic anomalies occur in the region where $x$ varies most quickly. }\label{fig: TK density extrema}
\end{figure*}
As discussed in the main text, the model we propose contains, for certain parameter values, a liquid-liquid phase transition within the equilibrium liquid free energy function. We discussed previously the interplay of temperature regimes, where the glass transition temperature $T_g$ should be compared with the temperature of the liquid-liquid critical point $T_c^{LL}$. Depending on the interplay of $T_g$ and $T_c^{LL}$, a variety of nucleation mechanisms may be relevant in the resulting material. In section \ref{ssec: Tg_crit}, we discussed a material where a liquid-liquid phase transition occurs only in the supercooled regime, i.e. $T_c^{LL}<T_g$ and pointed out that, while the liquid-liquid phase transition would in principle occur within a fully equilibrated sample, the phase transition is obscured by the glass transition for experiments on a human timescale. In fact, the model also permits that a material may have a putative liquid-liquid phase transition only below $T_K$. We examine one such material in figure \ref{fig: TK density extrema}. 

The parameters for this material are $\epsilon_b/\epsilon_{uu} = 7$, $\epsilon_{ub}/\epsilon_{uu} = \epsilon_{bb}/\epsilon_{uu} = 1$, $d_b/d_u = 1.2$. Notably, there is no frustration between bonded and nonbonded local structures encoded in the $\epsilon_{XY}$ values. Instead, a slight frustration coming from the repulsive forces is sufficient for a phase transition at low temperatures.

While mathematically the equilibrium liquid free energy for this material features a liquid-liquid phase transition, the phase transition would take place entirely below $T_K$, and thus no liquid-liquid phase transition would occur in the equilibrium liquid. Nevertheless, the water-like anomalies which result from the putative (actually avoided) liquid-liquid phase transition remain, even at equilibrium temperatures well above $T_g$. This scenario, where a putative liquid-liquid phase transition is obscured by a Kauzmann transition, would provide a singularity-free alternative to explaining water-like anomalies. Other singularity-free scenarios \cite{sastry_singularity-free_1996} have been proposed before.

\subsection{Extending the model to crystallization using the self-consistent phonon method}

\begin{figure*}
    \centering
    \includegraphics[width=\linewidth]{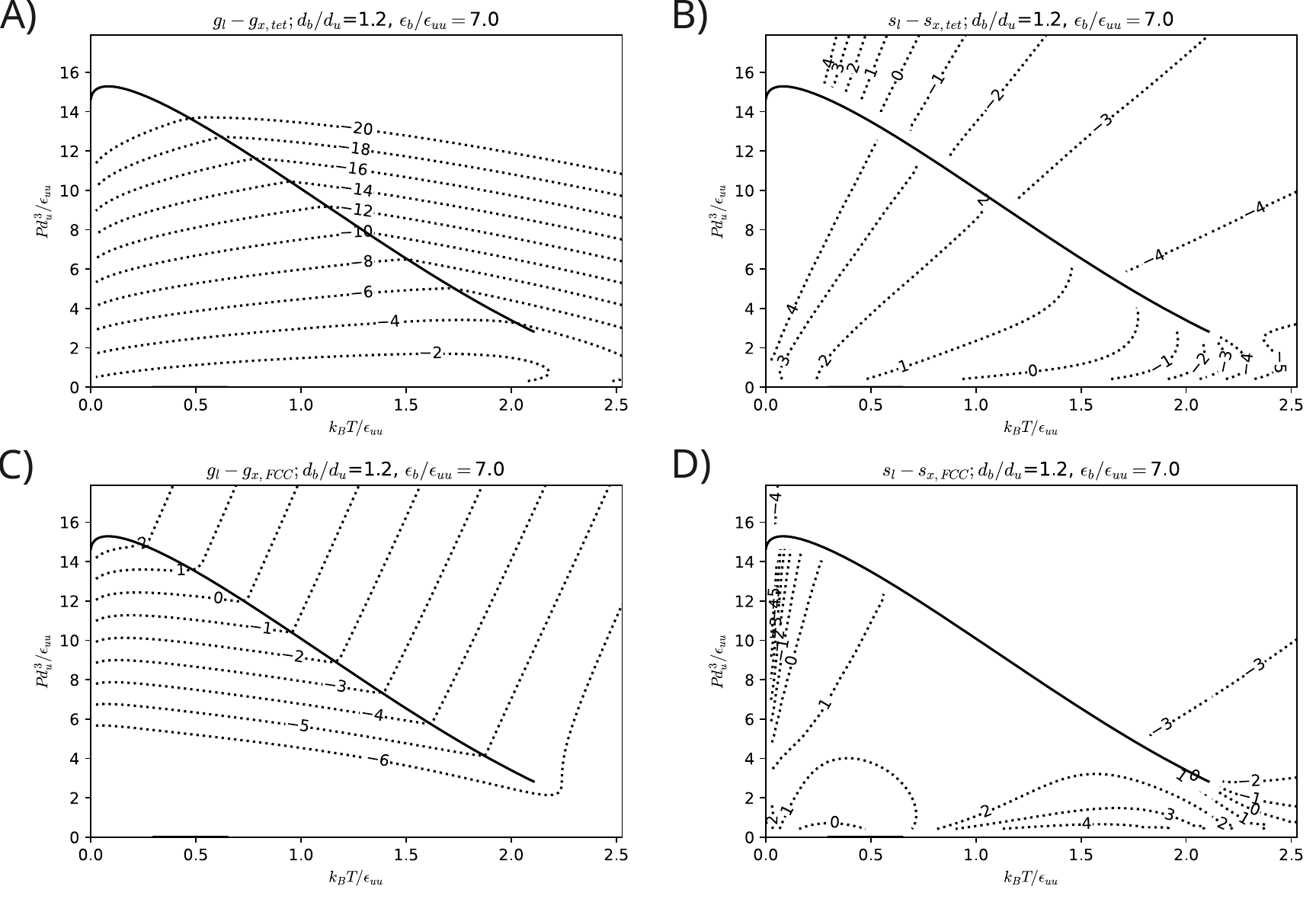}

    \caption{Here, we show the results of applying self-consistent phonon theory to two realistic crystal structures which are favorable for particles of our model. We investigate (a, b) a low density crystal characteristic of ordinary hexagonal water ice, where all particles may bond in perfect tetrahedrons. Additionally, we investigate (c, d) a face-centered cubic or hexagonal close packed lattice (either are equivalent under our approximations). The pair correlation function $g(r)$ is calculated as a sum of Dirac delta functions located at the pair distances of the corresponding crystal lattices. In (a) and (c), we show the resulting Gibbs free energy per particle, relative to the Gibbs free energy per particle of the liquid phase. At high pressures, the close packed, non-bonding crystal is ultimately stable. For these parameters, the low density crystal is never ultimately stable, as it can not attain a high enough density to overcome the attractive forces which stabilize the liquid. In (b) and (d), we show the entropy per particle of the crystals relative to the liquid.}\label{fig: Xtal}
\end{figure*}

Self-consistent phonon theory was originally developed to study the crystal melting transition of hard spheres \cite{fixman_highly_1969}. We can use it for this class of models by supplying the pair correlation function $g_x(\bm{R})$ corresponding to a particular crystal structure. We consider only two possible crystal lattices, a close packed face-centered cubic lattice made up of entirely nonbonding particles, and a hexagonal ice-like lattice made up of bonding particles. In both cases, the chosen crystal structure minimizes the energy of the configuration based on either the Lennard Jones attraction or through the bonding. We use a $g_x(\bm{R})$ based on the neighbor distances and number of neighbors of the two lattices, and sum Dirac delta functions at each neighbor distance. The vibration tensor $\alpha$ is inferred from $\alpha = \frac{\rho}{6}\int d\bm{R} g_x(\bm{R}) \nabla^2 V^{eff}(R)$, just as was calculated for the glassy states. We then vary the crystal density $\rho$, and thus the crystal lattice spacing $d_x\sim\rho^{-1/3}$, to determine the pressure.

In figure \ref{fig: Xtal}, we show the excess Gibbs free energy per particle (a, c) and the excess entropy per particle (b, d) of the liquid compared to these two crystals, a tetrahedral ice-like crystal (a-b) and a close packed face-centered cubic (c-d). The results for the dense FCC crystal are as to be expected: at low temperatures or high pressures, the liquid freezes, and the crystal has a lower entropy than the liquid. The results for the low density ice-like crystal are less water-like. Unlike in real water and ice, within our model, the liquid  has a lower enthalpy than the ice crystal. This is because despite having nearly four bonds per particle intact in the low density liquid, we have allowed additional particles within the nearest neighbor shell of the liquid. This reflects the fact that we have used an unrealistic $g(\bm{R})$ for the low density liquid, and in real low density water, $g(\bm{R})$ is highly anisotropic due to the bond angle constraints. In principle, there are also complex crystal structures with higher packing fraction but intact bonds, one of which is the likely ultimate ground state of our bonding particles.

\subsection{Nucleation in the proximity of both the critical point and the glass transition}

\begin{figure*}
    \centering
    \includegraphics[width=\linewidth]{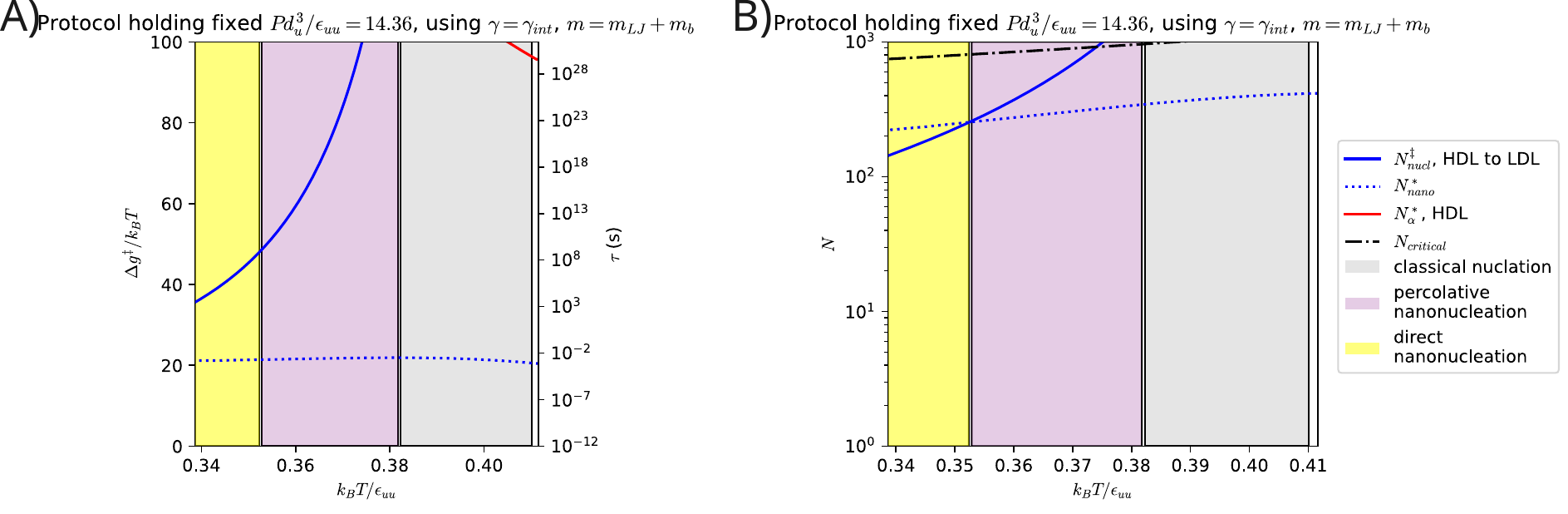}

    \caption{Here, we show the typical behavior of nucleation kinetics in the vicinity of both $T_g$ and $T_c^{LL}$. Nucleation rates are reported based on the timescale for nucleation to occur in a particular microscopic region, where for bulk nucleation, this region is a given particle and for nanonucleation, this region is the size of the cooperatively rearranging region. We use the same microscopic parameters as in section \ref{fig: Tg_crit}, resulting in a $T_g$ which is just above $T_c^{LL}$. Note that because the nucleation is, by necessity, occurring below $T_g$, the rate of glassy rearrangements and of bulk nucleation is extremely slow (and in fact, the latter is out of the scale of the current axes). However, nanonucleation still occurs at a relatively fast rate. Note however that the length scale of critical fluctuations is relatively large compared to the critical nucleus size for nanonucleation, and significant interface softening is to be expected in the nucleation process.}\label{fig: Tg_crit_nucleation}
\end{figure*}
When $T_g\gtrsim T_c^{LL}$, nucleation from one liquid to another is significantly effected by both critical fluctuations and glassy slowing down. While a full theory of ordinary critical fluctuations coupled to a random first-order phase transition is out of the scope of this paper, we have included in figure \ref{fig: Tg_crit_nucleation} the simplest picture of such a process. Here, we use the same parameters as section \ref{ssec: Tg_crit}, such that $T_g\approx T_c^{LL}$. Bulk nucleation does not occur on human timescales due to the dynamic prefactor from the glassy dynamics. However, nanonucleation still occurs on relatively fast timescales, and liquid-liquid nanonucleation may be observed even if $T_c^{LL}$ is below $T_g$. We stress again however, that the length scale of the critical fluctuations is larger than the length scale of the critical nucleus of nanonucleation, and thus this picture must be modified to account for the softening of the liquid-liquid droplet interface.

\subsection{Phase diagrams using silica-like parameters}
\begin{figure*}
    \centering
    \includegraphics[width=\linewidth]{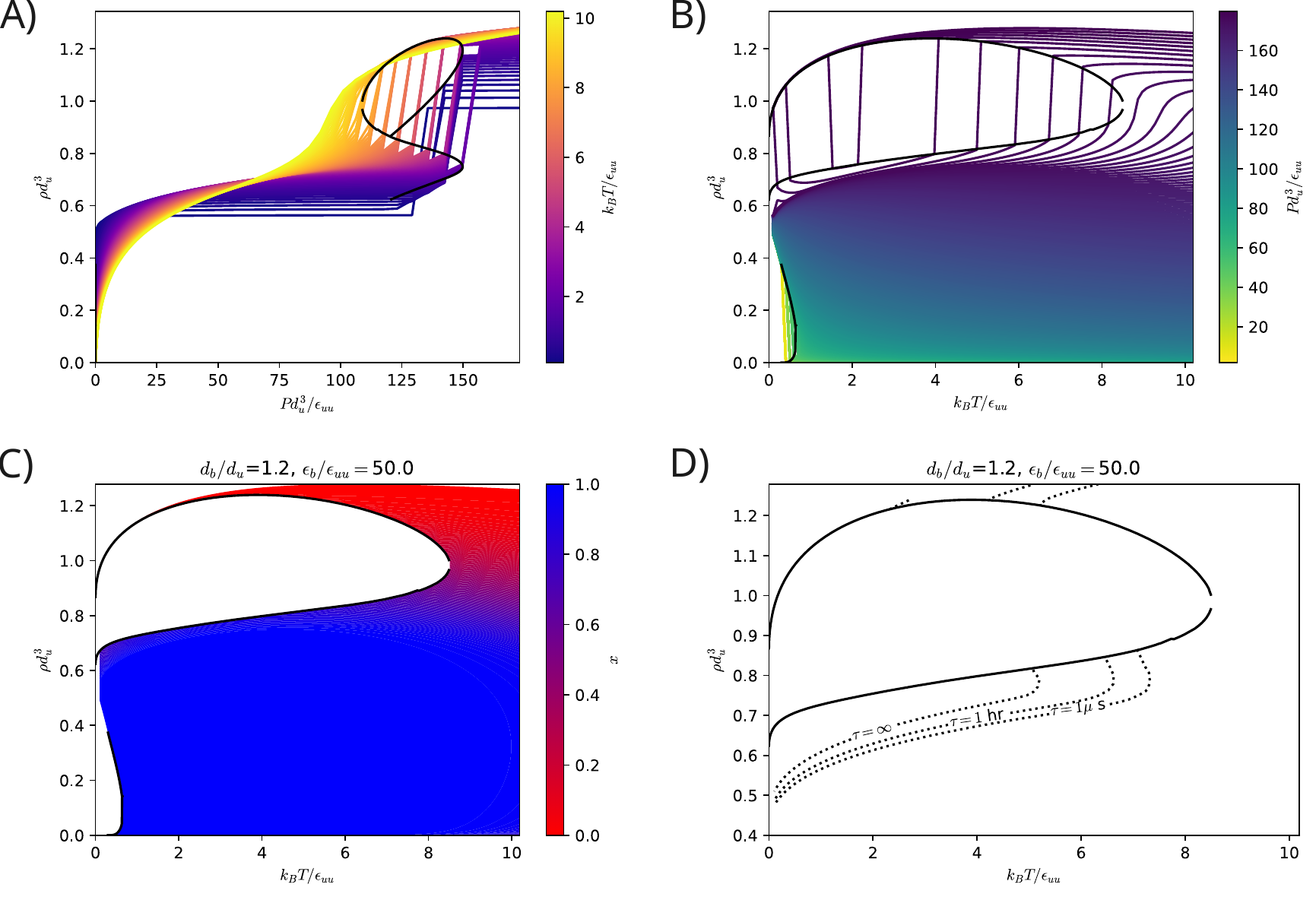}

    \caption{Here, we summarize the behavior of the model when $\epsilon_{ub}/\epsilon_{uu} = -1$ and $\epsilon_b/\epsilon_{uu} = 50$. This results in a liquid-liquid critical point which is modestly above the glass transition, as is the case for silica. Additionally, $\epsilon_b/\epsilon_{uu}$ reproduces the compositional dependence of fragility in a broad range of alkali silicates, as we showed previously \cite{brown_bonding_2025}. (a) $\rho$ vs $P$ isotherms, where the color indicates the temperature. There is clear evidence of a liquid-liquid transition. A liquid-gas transition is also present, but only for very low densities. (b) $\rho$ vs T isobars, where the color indicates the pressure. Note here that the liquid-gas transition is clearly visible in the lower left. (c) Fraction of particles which are non-bonding, $x$, plotted as a function of $\rho$ and $T$, where the color indicates the $x$ value. Except at and above temperatures comparable to the critical temperature, $x$ is nearly exactly 0 (solid red) or 1 (solid blue). (d) Lines of constant configurational entropy, corresponding to the string crossover at $s_c=1.28 k_B$, the laboratory glass transition at $s_c = 0.79 k_B$, and the Kauzmann transition at $s_c = 0$.}
    \label{fig: SiO2}
\end{figure*}

By tuning the strength of the bonds within our model to much higher values, we can also produce a phase diagram resembling that of silica. We use a bond strength $\epsilon_b/\epsilon_{uu} = 50$, which we found previously \cite{brown_bonding_2025} to be appropriate for describing alkali silicates in a similar model. Tuning to such a high bond strength does not qualitatively change the thermodynamics or kinetics of the liquid-liquid phase transition. However, increasing the bond strength does shift the phase transition to substantially higher pressures, when measured in microscopic units, reflecting the higher pressure needed to overcome the greater energetic cost of breaking bonds.

\end{document}